\documentclass[final,3p,times]{elsarticle}

\usepackage{amsmath,hyperref}
\usepackage{lineno}

\usepackage{epstopdf}
\journal{Journal}

\usepackage{lineno,hyperref}
\usepackage{graphicx}
\usepackage{dcolumn}
\usepackage{bm}
\usepackage{epsfig}
\usepackage{booktabs}
\usepackage{subfigure}
\usepackage{graphics}
\usepackage{amssymb}
\usepackage{amsmath}
\usepackage{array}
\usepackage{color}
\usepackage{booktabs}
\usepackage{multirow}
\usepackage{caption}
\usepackage{chngpage}
\usepackage{subfigure}
\usepackage{mathrsfs,gensymb,float}

\biboptions{numbers,sort&compress}
\modulolinenumbers[1]
\begin{document}

\captionsetup[figure]{labelfont={bf},name={Fig.},labelsep=period}        

\begin{frontmatter}
	
	\title{A diffuse-interface method for the containerless freezing of three-phase flows in complex geometries}
	

	
	
	
	
	

	\author[1]{Jiangxu Huang}
	\author[1]{Chengjie Zhan}
	\author[1,3,4,5]{Zhenhua Chai\corref{mycorrespondingauthor}}	
	\ead{hustczh@hust.edu.cn}
	\cortext[mycorrespondingauthor]{Corresponding author}
	\author[1]{Changsheng Huang}

	\address[1]{ School of Mathematics and Statistics, Huazhong University of Science and Technology, Wuhan 430074, China}
	\address[3]{Institute of Interdisciplinary Research for Mathematics and Applied Science,Huazhong University of Science and Technology, Wuhan 430074, China}	
	\address[4]{Hubei Key Laboratory of Engineering Modeling and Scientific Computing, Huazhong University of Science and Technology, Wuhan 430074, China}	
	
	\address[5]{The State Key Laboratory of Intelligent Manufacturing Equipment and Technology, Huazhong University of Science and Technology, Wuhan 430074, China}


\begin{abstract}

In this work, we first propose a diffuse-interface model for the freezing processes of three-phase flows in complex geometries, and the core of the model to intergratge the Navier-Stokes equations for fluid flows, a modified phase-field equation for gas-liquid interfaces, and an enthalpy approach for solid-liquid phase-change processes in a unified diffuse-interface framework. The volume expansion or shrinkage of the liquid phase caused by the density change during the phase-change process is considered by introducing a mass source term into the continuity equation. The wettability effect in such a gas-liquid-solid multiphase system is also included in the phase-field free energy, thereby avoiding the direct discretization of wetting boundary condition on the complex fluid-solid boundary. Then, we develop a mesoscopic lattice Boltzmann (LB) method to solve the diffuse-interface model for the freezing processes in multiphase systems, and test the accuracy and efficiency of the LB method through some benchmark problems, including the conduction-induced freezing in a semi-infinite space, the three-phase Stefan problem, the droplet solidification on the flat and curved surfaces. It is found that the numerical results are in good agreement with the experimental data and theoretical solutions. Finally, the LB method is further extended to study the freezing dynamics of multiphase flows in a fracture and porous medium, and the numerical results show that the developed method is efficient in the study of freezing processes of multiphase flows in complex geometries.


\end{abstract}

	
\end{frontmatter}

\section{Introduction}
Solidification or freezing is a common phenomenon in nature, and is also usually observed in many fields of applied science and engineering, including energy and environmental science \cite{AbhatSE1983, DuNRP2024}, geoscience \cite{DashRMP2006, McGuirePNAS2018}, material science \cite{DevilleScience2006}, construction engineering \cite{DevilleSpringer2017} and so on. To promote the development of technologies and to explore new applications in the aforementioned areas, it is crucial to gain a comprehensive understanding of the physical processes related to the freezing phenomena \cite{TiwariATE2023,HuerreARFM2024,WangPNAS2021}.

Over the past decades, many researchers have investigated the physical mechanisms of freezing by considering the fundamental problem of a single water droplet freezing on a cold substrate. For instance, Anderson et al. \cite{AndersonJCG1996} first proposed a theoretical model for the freezing on a flat surface by assuming the phase-change interface to be parallel to the cold surface, and found that the predicted freezing profiles are consistent with the experimental results. However, the predicted freezing time is different from the experimental data since the supercooling effect has been neglected in the theoretical model. Then Schultz et al. \cite{SchultzIDCS2001} developed a modified model, and demonstrated that the tip formed during the solidification of the droplet results from a combination of volume expansion and the curvature of the phase boundary. Later, this modified model is further extended by Ajaev et al. \cite{AjaevJCP2003} to consider the overhanging droplets. In addition, Zhang et al. \cite{ZhangATE2017} considered the effects of supercooling and gravity in the mathematical model, and illustrated that there is a good agreement between theoretical and experimental results in terms of freezing curve and freezing time. Virozub et al. \cite{VirozubJCG2008} pointed out that the assumption of a parallel freezing front was unreasonable, especially in the later stage of freezing. For this reason, Marín et al. \cite{MarinPRL2014} and Schetnikov et al. \cite{SchetnikovAJP2015} adopted a curved phase interface model, approximating the freezing front as a spherical cap. The results show that the freezing front keeps a nearly constant angle during the motion of droplet interface, and the top of the droplet eventually forms a pointed shape under the effects of density expansion and constant angle at the tri-junction points. Recently, Zhang et al. \cite{ZhangATE2019} addressed the limitations of curved interface model \cite{MarinPRL2014} by incorporating the effects of subcooling and gravity into the spherical-cap ice-water interface model, and accurately predicted the evolution of the profile throughout the freezing process. Although the theoretical and experimental studies mentioned above have enriched our understanding on the freezing process of a droplet, theoretical works are usually restricted to simple physical conditions and a limited range of physical parameters, and it is also difficult for experimental approaches to characterize detailed information about the interior of frozen droplets.

With the rapid development of computing techniques and advanced computational algorithms, numerical simulation has become an important tool in the study of freezing problems. The freezing process in an ambient fluid involves the gas, liquid and solid phases, the dynamic evolution of phase interfaces, and the coupling of flow and heat transfer, which bring some significant challenges in the numerical simulation of solidification. Actually, the existing methods for the freezing problem follow a similar procedure, i.e., one interface-capturing approach is adopted for the gas-liquid system, and another one is used for the solid-liquid phase-change system. Generally speaking, there are two kinds of methods that can be used to describe the interfacial evolution, i.e., the sharp-interface method \cite{SussmanJCP2007} and the diffuse-interface method \cite{AndersonARFM1998,YueJFM2004,DingJCP2007}. The former treats the gas-liquid phase boundary as a zero-thickness interface, requiring an explicit jump boundary condition and the tracking of the interface position \cite{AndersonARFM1998,YueJFM2004,DingJCP2007}. In contrast, the latter treats the interface as a smooth transition region with a non-zero thickness, and simultaneously, the difficulty in directly implementing of the jump boundary condition on the complex interface can be overcome \cite{SussmanJCP2007}. We note that several numerical approaches have been proposed to study the droplet freezing process on a flat substrate \cite{ZhangPRE2020, HuangJCP2022, ZhangJCP2022, WangJCP2024, WangJCP2025, MohammadipourJFM2024, HuangPRE2024, HuangIJHMT2025, VuIJMF2015, ShetabivashJCP2020, LyuJCP2021, YeCAMC2024, ThirumalaisamyIJMF2023, WeiJCP2024}, including the diffuse-interface method, the sharp-interface method, and the hybrid diffuse-sharp interface method. In terms of the diffuse-interface methods, Zhang et al. \cite{ZhangPRE2020} developed an axisymmetric pseudopotential multi-phase LB method to simulate the freezing process of a sessile water droplet, and at the same time, an enthalpy approach is applied to track the solid-liquid phase-change interface. Huang et al. \cite{HuangJCP2022} presented a phase-field model to study thermo-gas-liquid-solid flows with liquid-solid phase-change. In their model, the solid-liquid interface and droplet interface were captured by the Cahn-Hilliard (CH) and Allen-Cahn equations, respectively. Based on a similar idea, Zhang et al. \cite{ZhangJCP2022} and Wang et al. \cite{WangJCP2024, WangJCP2025} developed some multiple phase-field models to study the three-phase freezing problems. Mohammadipour et al. \cite{MohammadipourJFM2024} and Huang et al. \cite{HuangPRE2024,HuangIJHMT2025} developed a coupled enthalpy and phase-field method to investigate the solidification processes of droplets in three-phase systems, where the volume changes of droplets during the freezing process have been considered. With respect to the sharp-interface method, Vu et al. \cite{VuIJMF2015} proposed a front-tracking method for the three-phase freezing problem. In this method, both the droplet interface and the freezing front are tracked by front-tracking method, and the volume change resulting from density difference is also considered. Shetabivash et al. \cite{ShetabivashJCP2020} developed a multiple level-set approach for modeling the containerless freezing process in a three-phase system, in which the gas-liquid and solid-liquid phase interfaces are captured by two level sets. For the hybrid diffuse-sharp interface method, Lyu et al. \cite{LyuJCP2021} proposed a hybrid volume-of-fluid and immersed-boundary method (VOF-IBM) for the simulation of freezing processes of liquid films and droplet, and the volume expansion due to the density difference between ice and liquid is simultaneously included. Ye et al. \cite{YeCAMC2024} developed an adaptive hybrid level-set and moment-of-fluid method to study the impact and solidification of water droplet on the flat surface. Thirumalaisamy et al. \cite{ThirumalaisamyIJMF2023} proposed a hybrid let-set and low-Mach enthalpy method to investigate the solidification and melting problems in the presence of gas phase. In their method, the interface between the phase-change materials and gas phase is described by the level-set method, while the solid-liquid phase-change interface is tracked by the enthalpy method. Wei et al. \cite{WeiJCP2024} presented a hybrid phase-field and VOF method for the three-dimensional binary solidification in the presence of a gas bubble, where the phase-field method and VOF method are adopted to capture solidification and gas-liquid interfaces.

From above discussion, although there are many works on the freezing processes , most of them are limited to simple geometric structures, and the study of the freezing problems with the complex geometries remains scarce. Recently, Vu et al. \cite{VuIJHMT2021} adopted an axisymmetric three-phase interface-tracking method to study the freezing process of droplets in a gas environment. Zhang et al. \cite{ZhangATE2024} developed a theoretical model for droplet freezing on a sphere under the assumption that the freezing front is parallel to the curved surface, and also considered the effects of undercooling and gravity. However, these works cannot be directly extended to study the freezing problems with complex geometries, e.g.,  freezing in the porous media. Additionally, the wetting phenomenon of water on the ice is a complex and still-debated topic, but it is crucial for understanding the macroscopic structure of capillary flows on ice \cite{HuerreARFM2024,DotanJAST2009,MeulerACSANI2010,SarsharCPS2013}. Up to now, there are only a few numerical works considering the moving contact line in gas-liquid-ice/gas-liquid-solid multi-phase systems \cite{WeiJCP2024}. Furthermore, the volume change resulting from the density difference during phase change plays an important role in determining the evolution of the interfaces among different phases (e.g., the formation of the droplet tip after freezing). Although some numerical methods have been used to capture the droplet tip structures \cite{MohammadipourJFM2024, HuangPRE2024, ShetabivashJCP2020, ThirumalaisamyIJMF2023, WeiJCP2024, LyuJCP2021}, most of them still struggle to accurately reproduce this feature \cite{ChaudharyETFS2014,HagiwaraJCG2017,YaoATE2018,WangPOF2021,ChangATE2023,ZhuIJHMT2024,DaiPRE2024,FullerJCIS2024}, which is also a challenging problem in the numerical study of droplet freezing.

To address aforementioned issues, we propose a diffuse-interface model for the freezing processes in complex geometries, and the volume change induced by density difference is also considered. In this model, the phase-field method is used to capture the interface between the liquid and the surrounding gas phases, and the wetting information is reformulated into the free energy, thus avoiding the direct discretization of wetting boundary condition imposed on the complex boundary \cite{ZhanJCP2024}. In addition, compared to some traditional strategies coupling the interface-capturing methods with the temperature equations \cite{HuangJCP2022,ZhangJCP2022,WangJCP2024,WangJCP2025,VuIJMF2015,ShetabivashJCP2020,LyuJCP2021,YeCAMC2024,WeiJCP2024}, the present coupled phase-field and enthalpy method can also capture the motion of interface and the distribution of temperature field in a simple way. These two distinct features make the present method much simpler and more efficient in the study of the freezing problems with complex geometries. We would also like to point out that although the phase-field method can be used to describe solid-liquid phase-change problems and capture complex interface morphologies \cite{HuangJCP2022,ZhangJCP2022,WangJCP2024,WangJCP2025,WeiJCP2024}, it usually needs an initial interface definition for droplet freezing, which may not be consistent with the spontaneous nucleation in actual solidification processes. On the contrary, the enthalpy method does not require such an initial interface, enabling it to be more suitable for the nucleation and other phase-change problems. The rest of this paper is organized as follows. In Section \ref{sec2}, we present the physical problem of freezing in a complex geometry and the diffuse-interface model for such a complex problem, followed by the developed LB method in Section \ref{sec3}. In Section \ref{sec4}, some numerical experiments are performed to test the present LB method. Finally, some conclusions are given in Section \ref{sec5}.

\section{Physical problem and diffuse-interface model}
\label{sec2}
\begin{figure}[H]
	\centering
	\includegraphics[scale=0.45]{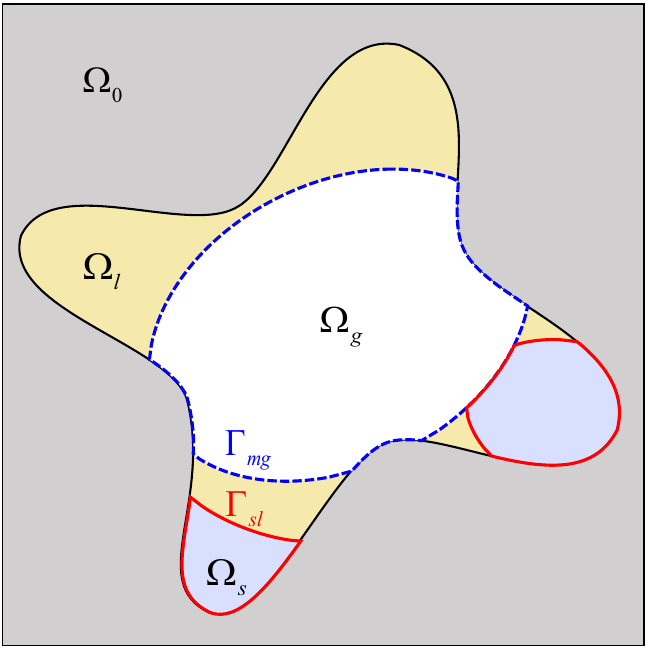} 
	\put(-155 ,132){(\textit{a})}
	\quad
	\quad
	\quad
	\includegraphics[scale=0.45]{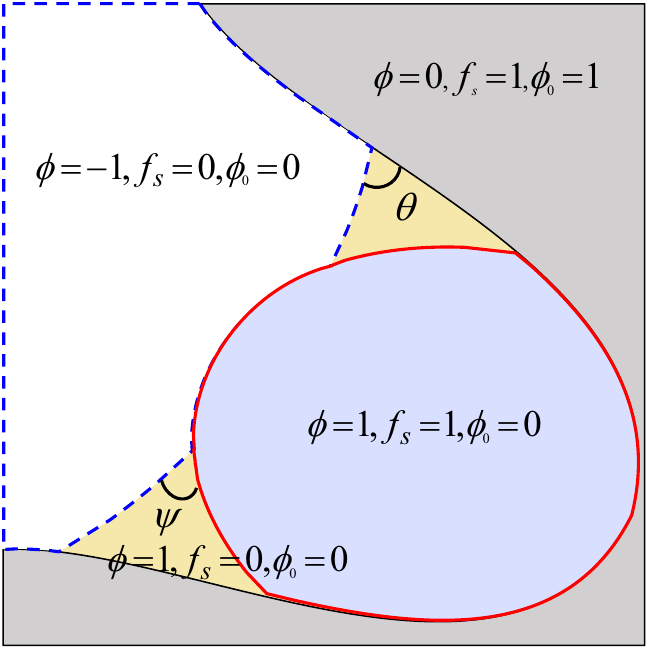} 
	\put(-155,132){(\textit{b})}
	\caption{ Schematic of three-phase freezing process in a complex geometry. (a) The gas-liquid-solid system with complex interfaces: the regions denoted by $\Omega_g$ (white region, $\phi=-1 \cap f_s=0 \cap \phi_0=0$), $\Omega_l$ (yellow region, $\phi=1 \cap f_s=0 \cap \phi_0=0$), and $\Omega_s$ (blue region, $\phi=1 \cap f_s=1 \cap \phi_0=0$) are filled with gas, liquid and solid phases. The complex region occupied by another solid phase is denoted as $\Omega_0$ (grey region, $\phi=0 \cap f_s=1 \cap \phi_0=1$). $\Gamma_{sl}$ indicates the freezing front between the solid and liquid phases, and $\Gamma_{mg}$ represents the interface between the phase-change material and gas phase. (b) The moving contact lines in gas–liquid–solid system, where $\theta$ is the contact angle between the liquid phase and pre-existing solid phase, while $\psi$ is the contact angle between the liquid phase and solid phase formed by freezing.}
	\label{fig1}
\end{figure}

In this section, we will present a diffuse-interface method for modeling three-phase freezing problems in complex geometries. The freezing process in the complex geometry is shown in Fig. \ref{fig1}(a), where the physical domain is $\Omega=\Omega_l \cup \Gamma_{sl} \cup \Omega_s \cup \Gamma_{mg} \cup \Omega_g \cup \Omega_0$, $\Gamma_{sl}$ denotes the solid-liquid interface or the freezing front, $ \Gamma_{mg} $ represents the interface between the solid-liquid mixture and gas phase, $\Omega_0$ is the complex domain, $\Omega_g$, $\Omega_l$ and $\Omega_s$ are the regions composed of the gas, liquid and solid phases, which are denoted by the subscripts $g$, $l$ and $s$. As the temperature is less than the freezing temperature $T_m$, the freezing process occurs from the liquid phase to the solid phase, and the freezing front $\Gamma_{sl}$ continuously advances. At the same time, the volume expansion or shrinkage caused by the solid-liquid density difference also results in the movement of $\Gamma_{mg}$. In addition, the wettability of solid surface is also needed to be considered in the gas-liquid-solid multiphase systems, which may involve two aspects: the first is the interaction between the liquid phase and the pre-existing solid phase, and the other is interaction between the newly formed solid phase and the liquid phase during the freezing process. It is worth noting that there is a general consensus that water should not completely wet ice \cite{HuerreARFM2024}, and there are many studies on the water-ice contact angle at both microscopic and macroscopic scales \cite{ElbaumPRL1991,MurataPNAS2016,KnightJCIS1967,KnightPM1971,ThievenazPRF2020}. However, due to the inability in achieving the perfect equilibrium, inadequate proof of purity, and the unknown shape of the ice-water interface at the tri-junction points, these measured contact angles may bring some uncertainties \cite{HuerreARFM2024}. Fig. \ref{fig1}(b) shows the complex interfaces between the liquid and solid phases during the freezing process, the contact angle $\theta$ characterizes the wettability of the pre-existing solid surface, while $\psi$ describes the contact angle between the liquid phase and the newly formed solid during freezing. Therefore, the freezing processes in the complex geometries include the dynamics of the phase interface and the dynamics of contact line. In this work, the modeling of complex freezing process is based on several assumptions:
\begin{itemize}
	\item The fluids are assumed to be Newtonian and immiscible. 
	\item The gas phase is not directly involved in the phase-change process, but acts as a moving and deformable boundary in the solidification process. 
	\item The liquid, solid and gas phases are considered as immiscible fluids, and there are no mass transfer among them. 
\end{itemize}
These assumptions provide the basis for applying the diffuse-interface method to describe the freezing process in a complex geometry. In the following, we will propose a mathematical model for containerless freezing processes in complex geometries, which includes a new phase-field equation for capturing the gas-liquid interface where the wettability effect of solid surface is also included, enthalpy-based energy equation for describing the solid-liquid interface, and Navier-Stokes equations for depicting the fluid flows.


In the freezing or solidification process, we only consider the density change near the freezing point, and the densities of the solid and liquid phases ($\rho_s$ and $\rho_l$) are assumed to be constants. In order to include the density change in the present model, some modifications are made to the continuity equation through neglecting the effect of gas phase. According to the principle of mass conservation before and after freezing \cite{HuangPRE2024}, we have
\begin{equation}
	\begin{aligned}
		& \frac{D}{D t}(M)=\frac{D}{D t}\left(M_l+M_s\right)=\frac{D}{D t}\left\{\int_{V(t)}\left[\rho_s f_s+\rho_l\left(1-f_s\right)\right] d V\right\}=0, \\
		\hookrightarrow & \int_{V(t)}\left\{\frac{\partial}{\partial t}\left(\rho_s f_s\right)+\nabla \cdot\left[\left(1-f_s\right) \rho_l \mathbf{u}_l\right]-\frac{\partial}{\partial t}\left(\rho_l f_s\right)\right\} d V=0, \\
		\hookrightarrow & \nabla \cdot\left[\left(1-f_s\right) \mathbf{u}_l\right]=\left(1-\frac{\rho_s}{\rho_l}\right) \frac{\partial f_s}{\partial t}, \\
		\hookrightarrow & \nabla \cdot \mathbf{u}=\left(1-\frac{\rho_s}{\rho_l}\right) \frac{\partial f_s}{\partial t}.
	\end{aligned}
	\label{eq1}
\end{equation}
where $\mathbf{u}$ and $f_s$ are the velocity and solid fraction. The source term $\dot{m}=(1-\frac{\rho_s}{\rho_l}) \frac{\partial f_s}{\partial t}$ on the right-hand side of the continuity equation describes the volume change during the freezing process, and it is influenced by the density ratio of solid to liquid phase.

The fluid movement is confined to the gas and liquid regions, and how to efficiently treat the fluid-solid boundary during the freezing process is a challenging problem. To overcome the difficulty in directly treating the no-slip boundary condition on the moving solid-fluid interface $\Gamma_{sl}$, several different sharp-interface approaches, such as the immersed boundary method or the embedded boundary method, have been proposed \cite{ShetabivashJCP2020,WeiJCP2024}. However, these methods are prone to encountering instability issues, particularly when forming the mushy region during freezing process \cite{WeiJCP2024}. In contrast, the diffuse-interface method with a non-zero thickness solid-fluid interface has been proved to be more robust in modeling fluid flow within such a mushy region. Specifically, this method can enforce the velocity to vanish in the solid region through introducing a penalty term into the momentum equation \cite{HuangPRE2024,HuangIJHMT2025},
\begin{equation}
	\frac{\partial (\rho \mathbf{u})}{\partial t}+\nabla \cdot(\rho \mathbf{u u})=-\nabla p+\nabla \cdot\left[\mu\left(\nabla \mathbf{u}+(\nabla \mathbf{u})^{\mathrm{T}}\right)\right]+\mathbf{F}_s+\mathbf{G} +  \rho \mathbf{f},
	\label{eq2}
\end{equation}
where $\rho$, $p$, $\mu$ are the density, pressure and dynamic viscosity, $\mathbf{G}$ is the body force. $ \mathbf{f} $ is the force generated by fluid-solid interaction to be given below, enforcing the no-slip boundary condition on the solid surface. $\mathbf{F}_s=\mu_\phi \nabla \phi$ is the surface tension force, and the chemical potential $\mu_\phi$ can be obtained by applying the variational derivative to the free energy functional \cite{ZhanJCP2024}.

Besides the Navier-Stokes equations for flow field, the interface between the liquid and gas phases can be captured by a modified CH equation,
\begin{equation}
	\frac{\partial \phi}{\partial t}+\nabla \cdot(\phi \mathbf{u})=\nabla \cdot \bar{M} \nabla \mu_\phi+\phi \nabla \cdot \mathbf{u},
	\label{eq3}
\end{equation}
where the order parameter $\phi$ is smoothly changed from 1 in liquid phase to -1 in gas phase, $\bar{M}$ is a positive constant named mobility. It should be noted that for incompressible fluid flows, the last term on the right hand side of above equation can be neglected, while it must be taken into account for the freezing process where the volume expansion or shrinkage would be caused by the density change. In addition, the wetting behavior of liquid on the complex boundary $\partial \Omega_0$ must also be considered since it has a significant influence on the interfacial dynamics. However, for the two-phase flows in complex geometries, the direct discretization of the wetting boundary condition on complex solid surface is very complicated, which may also break the mass conservation of the system. To avoid the direct implementation of wetting boundary condition, a new phase-field variable $\phi_0$ is introduced to label the complex geometry $\Omega_0$, and following the modeling approach for gas-liquid-solid three-phase flows \cite{ZhanJCP2024}, two phase-field variables $\phi_0$ and $\phi$ are used to distinguish the gas, liquid phases and solid phase in the complex region $\Omega_0$. By employing the mixed free energy of three-component fluid system proposed by Boyer and Lapuerta \cite{BoyerESAIM2006}, along with the Young's equation, we can obtain a new energy functional  $\mathcal{F}_\phi$ for the gas-liquid-solid multiphase system involving the wettability effect of solid surface,
\begin{equation}
	\begin{aligned}
		\mathcal{F}_\phi=\int_{\Omega}[\underbrace{\frac{3 \sigma}{4 \varepsilon}(1-\phi)^2(1+\phi)^2+\frac{3 \varepsilon \sigma}{16} |\nabla \phi |^2}_{\text {Standard free energy density for two-phase system}} & +\underbrace{\frac{9 \sigma}{2 \varepsilon} \phi_0{ }^2 \phi^2}_{\text {penalty term}}+\underbrace{\frac{3 \sigma \cos \theta}{\varepsilon} \phi_0 \phi\left(\phi^2+\phi_0^2-1\right)+\frac{3 \varepsilon \sigma \cos \theta}{8} \nabla \phi_0 \cdot \nabla \phi}_{\text {Wetting property of the solid surface $\Omega_0$}} \\
		& +\underbrace{\frac{9 \sigma}{2 \varepsilon} f_s{ }^2 \phi^2}_{\text {penalty term}}+\underbrace{\frac{3 \sigma \cos \psi}{\varepsilon} f_s \phi\left(\phi^2+f_s{ }^2-1\right)+\frac{3 \varepsilon \sigma \cos \psi}{8} \nabla f_s \cdot \nabla \phi}_{\text {Wetting property of the formed solid surface $\Omega_s$}}] d \Omega,
	\end{aligned}
	\label{eq44}
\end{equation}
where $\varepsilon$ is the interface thickness. The terms on the right-hand side of Eq. (\ref{eq44}) represent, from left to right, the standard free-energy density for two-phase flows, an additional term to reflect the penalty in the solid phase $\Omega_0$, the wettability of the solid surface $\Omega_0$, an additional term to reflect the penalty in the phase $\Omega_s$, and the wettability on the ice surface $\partial \Omega_s$.

\begin{figure}[H]
	\centering
	\includegraphics[width=0.4\textwidth]{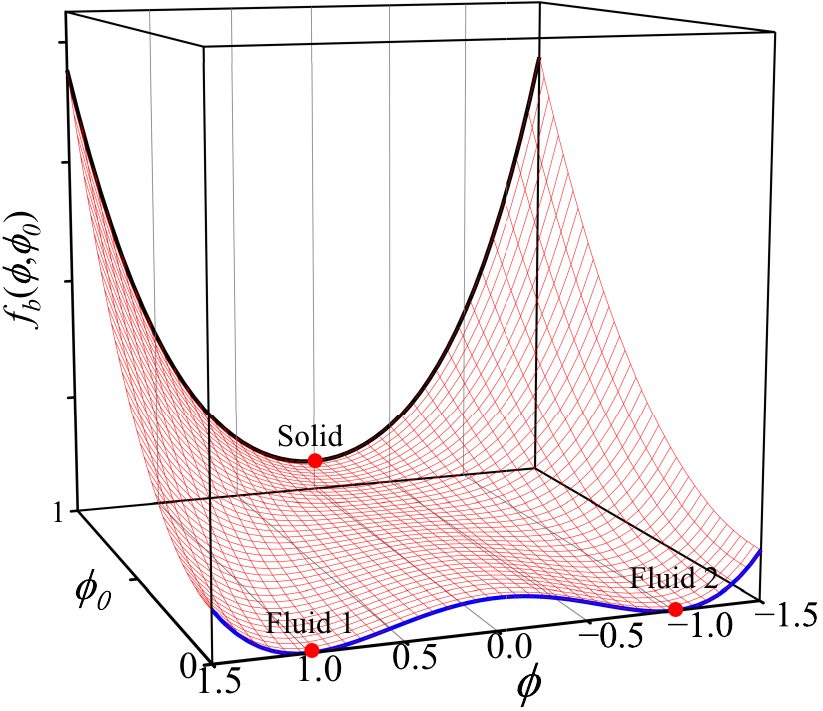} 
	\caption{ Distribution of bulk free energy density $f_b\left(\phi, \phi_0\right)=\frac{3 \sigma}{4 \varepsilon}(1-\phi)^2(1+\phi)^2+\frac{9 \sigma}{2 \varepsilon} \phi_0^2 \phi^2$ with $\sigma=0.02$ and $\varepsilon=3$.  }
	\label{fig1_add}
\end{figure}

To make a clear understanding on above energy functional, we can define a new bulk free energy density \cite{ZhanJCP2024},
\begin{equation}
	f_b\left(\phi, \phi_0\right)=\frac{3 \sigma}{4 \varepsilon}(1-\phi)^2(1+\phi)^2+\frac{9 \sigma}{2 \varepsilon} \phi_0^2 \phi^2,
\end{equation}
and plot it in Fig. \ref{fig1_add}. As shown in this figure, when $\phi_0=0$, corresponding to the fluid region, the bulk free energy density $f_b\left(\phi, \phi_0\right)$ reduces to the original double-well potential with two minima at $\phi= \pm 1$. However, when $\phi_0=1$, corresponding to the solid phase in $\Omega_0$, it transforms into a single-well potential with a minimum at $\phi=0$. The most distinct feature of this method is that the wettability can be included in the energy functional, and we do not need to implement the wetting boundary condition on complex solid surface. It is worth noting that the proposed free-energy functional would reduce to the standard free-energy functional of the binary fluid flows when $\phi_0=0$ and $f_s=0$. Taking the variational operator to the energy  functional $\mathcal{F}_\phi$, we have
\begin{equation}
	\begin{aligned}
		\mu_\phi=\frac{\delta \mathcal{F}_\phi}{\delta \phi}=\frac{3 \sigma}{\varepsilon} \phi(\phi-1)(\phi+1)-\frac{3 \varepsilon \sigma}{8} \nabla^2 \phi & +\frac{9 \sigma}{\varepsilon} \phi_0^2 \phi+\frac{3 \sigma \cos \theta}{\varepsilon} \phi_0\left(3 \phi^2+\phi_0^2-1\right)-\frac{3 \varepsilon \sigma \cos \theta}{8} \nabla^2 \phi_0 \\
		& +\frac{9 \sigma}{\varepsilon} f_s^2 \phi+\frac{3 \sigma \cos \psi}{\varepsilon} f_s\left(3 \phi^2+f_s^2-1\right)-\frac{3 \varepsilon \sigma \cos \psi}{8} \nabla^2 f_s .
	\end{aligned}
\end{equation}

Substituting above equation into Eq. (\ref{eq3}), we can obtain a new phase field equation for gas-liquid interface. Apart from the above hydrodynamic and phase-field equations, the energy equation based on enthalpy is adopted to describe the freezing process, which can be expressed as \cite{HuangPRE2024}
\begin{equation}
	\frac{\partial \left(\rho H\right)}{\partial t}+\nabla \cdot\left(\rho C_p T \mathbf{u}\right)=\nabla \cdot(\lambda \nabla T) + \rho C_p T \dot{m},
	\label{eq4}
\end{equation}
where $C_p$ is the specific heat capacity, $T$ is the temperature, $\lambda$ is the thermal conductivity, $H=C_p T+L f_l $ is the total enthalpy with $L$ being the latent heat. The last term on the right-hand side of Eq. (\ref{eq4}) is derived from $\nabla \cdot \mathbf{u}=\dot{m}$, and it can be ignored for the incompressible fluid flows. The temperature $T$ and the liquid phase fraction $f_l$ can be uniquely determined by the total enthalpy $\rho H$ \cite{ZhaoAML2020},
\begin{equation}
	f_l =\left\{\begin{array}{ll}
		0 & \rho H< \rho_s H_s \\
		\frac{\rho H - \rho_s H_s}{\rho_l H_l-\rho_s H_s} &\rho_s  H_s \leqslant \rho H \leqslant \rho_l H_l, \\
		1 & \rho H>\rho_l H_l
	\end{array} \quad T= \begin{cases} \frac{\rho H}{\rho_s  C_{p,s}}  & \rho H< \rho_s H_s \\
		T_s+\frac{\rho H-\rho_s H_s}{\rho_l H_l-\rho_s H_s}\left(T_l-T_s\right) & \rho_s H_s \leqslant \rho H \leqslant \rho_l H_l, \\
		T_l+ \frac{\rho H-\rho_l H_l}{\rho_l C_{p,l}}  & \rho H>\rho_l H_l\end{cases}\right.
	\label{Hfl}
\end{equation}
where $T_s$ and $T_l$ are the solidus and liquidus temperatures, respectively. $H_s$ and $H_l$ are the total enthalpies corresponding to the solidus and liquidus temperatures. It is worth noting that the present formulas for calculating $f_l$ and $T$ take into account the solid-liquid density difference, while it has been neglected in the previous work \cite{HuangJCP2022}, leading to an inconsistency with flow field \cite{ZhaoAML2020}, and additionally, Eq. (\ref{Hfl}) would reduce to the typical formula when $\rho_s = \rho_l $ \cite{ZhaoAML2020}. We would also like to pint out that the solid–liquid interface in the computational domain is implicitly tracked by using the solid fraction variable $f_s=1-f_l$ that is defined over the entire domain, $f_s=0$ and $f_s=1$ denote the fluid and solid phases, and $0<f_s<1$ represents the mushy zone.

The present phase-field model is composed of Eqs. (\ref{eq1}), (\ref{eq2}), (\ref{eq3}) and (\ref{eq4}), and it can be used to describe the three-phase freezing processes in complex geometries. The interfaces ($\Gamma_{sl}$ and $\Gamma_{mg}$) among different phases can be tracked simultaneously by the order parameter $\phi$ and solid fraction $f_s$, and the domains occupied by gas, liquid and solid phases ($\Omega_g$, $\Omega_l$, $\Omega_s$ and $\Omega_0$) can be identified by 
\begin{equation}
	\text { Phase domain }=\left\{\begin{array}{llllll}
		\Omega_g: & \phi=-1 & \& & f_s=0 & \& & \phi_0=0 \\
		\Omega_l: & \phi=1 & \& & f_s=0 & \& & \phi_0=0 \\
		\Omega_s: & \phi=1 & \& & f_s=1 & \& & \phi_0=0 \\
		\Omega_0: & \phi=0 & \& & f_s=1 & \& & \phi_0=1
	\end{array}\right..
\end{equation}
In this case, the physical property of system can be characterized by a simple linear function of the order parameter $\phi$, solid fraction $f_s$ and $\phi_0$,
\begin{equation}
	\begin{aligned}
		 \zeta=f_s \zeta_s+\left(1-f_s\right) \frac{1-\phi_0+\phi}{2} \zeta_l+\left(1-f_s\right)\left(1-\frac{1-\phi_0+\phi}{2}\right) \zeta_g, 
	\end{aligned}
\end{equation}
where the parameter $\zeta$ denotes a physical variable or coefficient, such as the density $\rho$, dynamic viscosity $\mu$, thermal conductivity $\lambda$, or specific heat capacity $C_p$.

\section{Numerical methods}
\label{sec3}

In this section, we develop a numerical method to solve the diffuse-interface model proposed in the previous section. The LB method, as a kinetic-theory based mesoscopic numerical approach, has some distinct particle-based properties and advantages in the study of multiphase flows \cite{ChenARFM1998, AidunARFM2010, KrugerBook2017}. In the following, will present a new LB method for the freezing of three-phase flows in complex geometries, which is composed of three LB models for flow, phase and temperature fields. To maintain the simplicity and computational efficiency of the LB method, we only consider the single-relaxation-time LB model, also known as the lattice BGK model \cite{QianEL1992},  for the fluid and phase fields, and the two-relaxation-time LB model for temperature field, which can be also easily extended to the advanced multiple-relaxation-time LB model \cite{ChaiPRE2020}.

\subsection{LB model for the flow field}

Unlike the original lattice BGK model for incompressible flows, now we introduce an additional mass source in the evolution equation to reflect volume change caused by the density difference during the freezing process \cite{HuangPRE2024,HuangIJHMT2025,YuanCMA2020},  

\begin{equation}
	f_i\left(\mathbf{x}+\mathbf{c}_i \Delta t, t+\Delta t\right)-f_i(\mathbf{x}, t)=-\frac{1}{\tau_f}\left[f_i(\mathbf{x}, t)-f_i^{\mathrm{eq}}(\mathbf{x}, t)\right]+\Delta t\left(1-\frac{1}{2 \tau_f}\right) F_i(\mathbf{x}, t),
\end{equation}
where $f_i(\mathbf{x}, t)$ is the distribution function at position $\mathbf{x}$ and time $t$, $\mathbf{c}_i$ is the discrete velocity, $\Delta t$ is the time step. $\tau_f=\nu /c_s^2 \Delta t + 0.5$ is the relaxation time for flow field with $\nu=\mu / \rho$ being the kinematic viscosity, $c_s = c /\sqrt 3$ is the sound speed. For the two-dimensional LB method, we adopt the nine-velocity (D2Q9) lattice model, the weight coefficients $\omega_i$ are given as $\omega_0=4/9$, $\omega_{1-4}=1/9$ and $\omega_{5-8}=1/36$, and the discrete velocities $\mathbf{c}_i$ are defined as
\begin{equation}
	\mathbf{c}_i= \begin{cases}(0,0), & i=0, \\ (\cos [(i-1) \pi / 2], \sin [(i-1) \pi / 2]) c, & i=1-4, \\ (\cos [(2 i-9) \pi / 4], \sin [(2 i-9) \pi / 4]) \sqrt{2} c, & i=5-8,\end{cases}
\end{equation}
where $c = \Delta x/\Delta t$ is the lattice speed with $\Delta x$ denoting the lattice spacing. For the three-dimensional LB method, one can consider the nineteen-velocity (D3Q19) lattice model, where the weight coefficients $\omega_i$ are set to be $\omega_0=1/3$, $\omega_{1-6}=1/18$ and $\omega_{7-18}=1/36$, and the discrete velocities are given by \cite{HuangCF2024}
\begin{equation}
	\mathbf{c}_i=c\left[\begin{array}{ccccccccccccccccccc}
		0 & 1 & -1 & 0 & 0 & 0 & 0 & 1 & -1 & 1 & -1 & 1 & -1 & 1 & -1 & 0 & 0 & 0 & 0 \\
		0 & 0 & 0 & 1 & -1 & 0 & 0 & 1 & -1 & -1 & 1 & 0 & 0 & 0 & 0 & 1 & -1 & 1 & -1 \\
		0 & 0 & 0 & 0 & 0 & 1 & -1 & 0 & 0 & 0 & 0 & 1 & -1 & -1 & 1 & 1 & -1 & -1 & 1
	\end{array}\right].
\end{equation}
To recover the Navier-Stokes equations correctly, the local equilibrium distribution function $f_i^{\mathrm{eq}}$ is defined by \cite{HuangPRE2024}
\begin{equation}
	f_i^{\mathrm{eq}}= \begin{cases}\frac{p}{c_s^2}\left(\omega_i-1\right)+\rho s_i(\mathbf{u}), & \mathrm{i}=0 \\ \frac{p}{c_s^2} \omega_i+\rho s_i(\mathbf{u}), & \mathrm{i} \neq 0\end{cases},
\end{equation}
where
\begin{equation}
	s_i(\mathbf{u})=\omega_i\left[\frac{\mathbf{c}_i \cdot \mathbf{u}}{c_s^2}+\frac{\left(\mathbf{c}_i \cdot \mathbf{u}\right)^2}{2 c_s^4}-\frac{\mathbf{u} \cdot \mathbf{u}}{2 c_s^2}\right].
\end{equation}
The force term $F_i$ is given by  \cite{HuangPRE2024}
\begin{equation}
	F_i=\omega_i\left[S+\frac{\mathbf{c}_i \cdot(\mathbf{F}+\rho \mathbf{f})}{c_s^2}+\frac{(\mathbf{u} \tilde{\mathbf{F}}+\tilde{\mathbf{F}} \mathbf{u}):\left(\mathbf{c}_i \mathbf{c}_i-c_s^2 \mathbf{I}\right)}{2 c_s^4}\right],
\end{equation}
where $S=\rho \dot{m}+\mathbf{u} \cdot \nabla \rho$, $\tilde{\mathbf{F}}=\mathbf{F}-\nabla(p-c_s^2\rho) +c_s^2 S$ and $\mathbf{F}=\mathbf{F}_{\mathrm{s}}+\mathbf{G}$ is the total force. The macroscopic velocity $\mathbf{u}$ and pressure $p$ can be evaluated by  \cite{HuangPRE2024}
\begin{subequations}
	\begin{equation}
		\rho \mathbf{u}^*=\sum \mathbf{c}_i f_i+\frac{\Delta t}{2} \mathbf{F},
	\end{equation}
	\begin{equation}
		\mathbf{u}=\mathbf{u}^*+\frac{\Delta t}{2} \mathbf{f},
	\end{equation}
	\begin{equation}
		p=\frac{c_s^2}{\left(1-\omega_0\right)}\left[\sum_{i \neq 0} f_i+\frac{\Delta t}{2} S+(\tau_f-0.5) \Delta t F_0+\rho s_0(\mathbf{u})\right],
	\end{equation}
\end{subequations}
where $\mathbf{u}^*$ is the velocity without considering the solid-liquid interaction, the force $\mathbf{f}=f_s \left(\mathbf{u}_s-\mathbf{u}^*\right) / \Delta t$ can be used to include fluid-solid interaction with $\mathbf{u}_s$ being the solid-phase velocity \cite{LiuCICP2022, LiuCF2024, ZhanCICP2023, ZhanPD2024, HuangPRE2024, HuangIJHMT2025}. We note that this approach has been recently applied to study various problems involving complex fluid-solid interfaces, including particulate flows \cite{LiuCICP2022, LiuCF2024}, dendrite growth \cite{ZhanCICP2023}, phase change and fluid flows in complex geometries \cite{HuangPRE2024, HuangIJHMT2025, ZhanPD2024}.

\subsection{LB model for the phase field}
For the modified CH equation with an extra source term  for the phase field, the LB evolution equation can be written as \cite{ZhanJCP2024}
\begin{equation}
	g_i\left(\mathbf{x}+\mathbf{c}_i \Delta t, t+\Delta t\right)-g_i(\mathbf{x}, t)=-\frac{1}{\tau_g}\left[g_i(\mathbf{x}, t)-g_i^{\mathrm{eq}}(\mathbf{x}, t)\right]+\left(1-\frac{1}{2 \tau_g}\right) \Delta t G_i(\mathbf{x}, t).
\end{equation}
Here local equilibrium distribution function $g_i^{\mathrm{eq}}$ and the distribution function of  source term $G_i$ are given by
\begin{equation}
	g_i^{\mathrm{eq}}(\mathbf{x}, t)= \begin{cases}\phi+\left(\omega_i-1\right) \mu_\phi, & i=0 \\ \omega_i  \mu_\phi+\omega_i \frac{\mathbf{c}_i \cdot \phi \mathbf{u}}{c_s^2}, & i \neq 0\end{cases},
\end{equation}
\begin{equation}
	G_i=\frac{\omega_i \mathbf{c}_i \cdot \partial_t(\phi \mathbf{u})}{c_s^2}+\omega_i \phi \nabla \cdot \mathbf{u}.
\end{equation}
In addition, through the Chapman-Enskog expansion \cite{ChaiPRE2020}, one can determine the relation between the mobility $\bar{M}$ and relaxation time $\tau_g$,
\begin{equation}
	\bar{M}=\eta c_s^2\left(\tau_g-0.5\right) \delta t,
\end{equation}
and the order parameter is calculated by
\begin{equation}
	\phi=\sum_i g_i+\frac{\Delta t}{2} \phi \nabla \cdot \mathbf{u}.
\end{equation}

\subsection{LB model for the temperature field}
To avoid numerical diffusion across solid-liquid interface and consider computational accuracy and efficiency \cite{HuangJCP2015}, we adopt the two-relaxation-time LB model to solve the energy equation (\ref{eq4}). The evolution equation of the LB model for temperature field reads \cite{LuIJTS2019}
\begin{equation}
	h_{\mathrm{i}}\left(\mathbf{x}+\mathbf{e}_i \Delta t, t+\Delta t\right)=h_{\mathrm{i}}(\mathbf{x}, t)-\frac{1}{\tau_h^\mathrm{a}}\left[h_{i}^{\mathrm{a}}(\mathbf{x}, t)-h_{i}^{\mathrm{aeq}}(\mathbf{x}, t)\right]-\frac{1}{\tau_h^{\mathrm{s}}}\left[h_{i}^{\mathrm{eq}}(\mathbf{x}, t)-h_{i}^{\mathrm{seq}}(\mathbf{x}, t)\right]+\left(1-\frac{1}{2 \tau_h^{\mathrm{a}}}\right) \Delta t H_i(\mathbf{x}, t),
	\label{eq22}
\end{equation}
where $h_{i}^{\mathrm{s}}(\mathbf{x}, t)$ and $h_{i}^{\mathrm{a}}(\mathbf{x}, t)$ are the symmetric and anti-symmetric parts of distribution function $h_{i}(\mathbf{x}, t)$ for the total enthalpy $\rho H$, $\tau_h^{\mathrm{s}}$ and $\tau_h^{\mathrm{a}}$ are the symmetric and anti-symmetric relaxation times, $ h_{i}^{\mathrm{seq}}(\mathbf{x}, t) $ and $ h_{i}^{\mathrm{aeq}}(\mathbf{x}, t) $ are the symmetric and anti-symmetric parts of the equilibrium distribution function. The expressions of $h_{i}^{\mathrm{s}}(\mathbf{x}, t)$, $h_{i}^{\mathrm{a}}(\mathbf{x}, t)$, $ h_{i}^{\mathrm{seq}}(\mathbf{x}, t) $ and $ h_{i}^{\mathrm{aeq}}(\mathbf{x}, t) $ are given by \cite{LuIJTS2019}
\begin{equation}
	h_{i}^{\mathrm{s}}=\frac{h_i+h_{\bar{i}}}{2}, h_{i}^{\mathrm{a}}=\frac{h_i-h_{\bar{i}}}{2}, h_{i}^{\mathrm{seq}}=\frac{h_i^{\mathrm{eq}}+h_{\bar{i}}^{\mathrm{eq}}}{2}, h_{i}^{\mathrm{aeq}}=\frac{h_i^{\mathrm{eq}}-h_{\bar{i}}^{\mathrm{eq}}}{2},
\end{equation}
where $\bar{i}$ represents the opposite direction of $i$, i.e.,$\mathbf{c}_{\bar{i}}=-\mathbf{c}_i$. The equilibrium distribution function for the total enthalpy can be defined as \cite{LuIJTS2019}
\begin{equation}
	h_i^{\mathrm{eq}}= \begin{cases}\rho H-\rho C_{p, \mathrm { ref }} T+\omega_i \rho C_p T\left(\frac{\rho_{\mathrm { ref }} C_{p, \mathrm { ref }}}{\rho C_p}-\frac{\mathbf{I}: \mathbf{u u}}{2 c_s^2}\right), & i=0 \\ \omega_i \rho C_p T\left[\frac{\rho_{\mathrm { ref }} C_{p . \mathrm{ref}}}{\rho C_p}+\frac{\mathbf{c}_i \cdot \mathbf{u}}{c_s^2}+\frac{\left(\mathbf{c}_i \mathbf{c}_i-c_s^2 \mathbf{I}\right): \mathbf{u u}}{2 c_s^4}\right], & i \neq 0 \end{cases}.
\end{equation}
where $C_{p, \mathrm{ ref }}=2 C_{p, s} C_{p, l} /\left(C_{p, s}+C_{p, l}\right)$ is the capacity specific heat, and $\rho_{\mathrm { ref }}$ is the reference density. In order to recover the energy equation (\ref{eq4}), a source term $	H_i(\mathbf{x}, t)=\rho C_p T \nabla \cdot \mathbf{u}$ is introduced into the evolution equation, the thermal conductivity should be related to the anti-symmetric relaxation time,
\begin{equation}
	\frac{\lambda}{\rho c_{p, \mathrm { ref }}}=\left(\tau_h^\mathrm{a}-0.5\right) c_s^2 \Delta t,
\end{equation}
As discussed in Ref. \cite{LuIJTS2019}, the present two-relaxation time LB model can reduce the numerical diffusion across the solid-liquid interface once the following relation is satisfied, 
\begin{equation}
	 \frac{1}{\tau_h^\mathrm{s}}+\frac{1}{\tau_h^\mathrm{a}}=2 .
\end{equation}
In addition, the total enthalpy is calculated by \cite{HuangPRE2024}
\begin{equation}
	\rho H=\sum_{i=0} h_i+\frac{\Delta t}{2}  \rho C_p T \dot{m}.
\end{equation}
When the total mixed enthalpy is known, we can update the temperature and liquid fraction with Eq. (\ref{Hfl}).

\section{Numerical results and discussion}
\label{sec4}

In this section, we will first perform some simulations of conduction-induced freezing in a semi-infinite space, the three-phase Stefan problem, the droplet freezing on flat and curved surfaces, and conduct comparisons with some available analytical, numerical and experimental data to test the accuracy of LB method. Then the LB method is applied to investigate the freezing processes in a rough fracture and unsaturated porous medium to demonstrate the capacity of present method in the study of the freezing dynamics within complex geometries. Before performing any numerical tests, we introduce the following dimensionless parameters to characterize the dynamics of the freezing process, including the Stefan number ($\mathit{Ste}$), the Prandtl number ($Pr$) and the Fourier number ($Fo$),
\begin{equation}
	\mathit{Ste}=\frac{C_{p, l}\left(T_m-T_w\right)}{L}, \mathit{Pr}=\frac{v}{\alpha}, F o=\frac{\lambda_l t}{\rho_l C_{p, l} R_0^2},
\end{equation}
where $\mathit{Ste}$ is defined as the ratio of sensible and latent heat, $R_0$ is the reference length, $\mathit{Pr}$ represents the ratio of momentum to heat diffusion, and $Fo$ is the dimensionless time.

\subsection{Conduction freezing in a semi-infinite space}
\begin{figure}[H]
	\centering
	\includegraphics[scale=0.5]{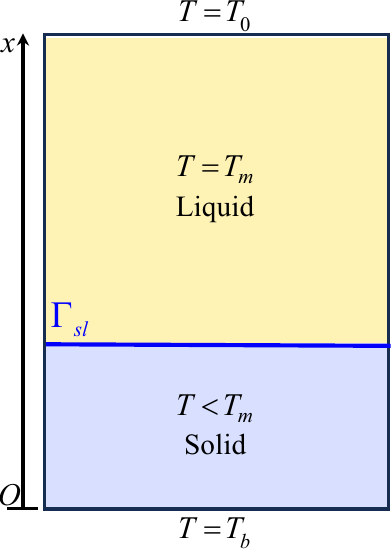} 
	\label{fig2a}
	\put(-110 ,120){(\textit{a})}
	\quad
	\includegraphics[scale=0.85]{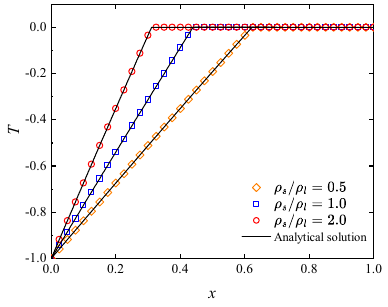} 
	\label{fig2b}
	\put(-165,120){(\textit{b})}
	\includegraphics[scale=0.85]{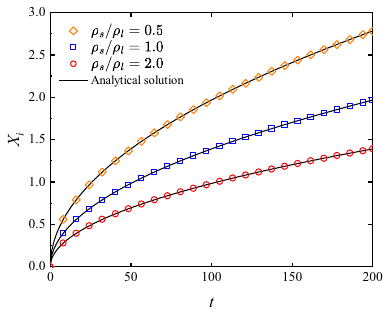} 
	\label{fig2c}
	\put(-165,120){(\textit{c})}
	\caption{ Schematic of the one-phase freezing problem in the semi-infinite space (a). Comparisons of the temperature distribution (b) and the solid-liquid interface evolution (c) between the numerical results and analytical solutions under different solid-liquid density ratios $\rho_s/ \rho_l$.}
	\label{fig2}
\end{figure}

The thermal-conduction induced freezing in a semi-infinite space is a classical heat transfer problem, and is usually used to test the accuracy of numerical method. In this part, we first simulate the one-phase freezing by conduction to test total enthalpy based LB model. Here one-phase means that the freezing occurs at a constant temperature and the liquid phase stays steadily at a freezing temperature $T_m$. The schematic of the problem is shown in Fig. \ref{fig2}(a), where the entire region is initially filled with the liquid phase at a uniform temperature $T_0$ ($T_0=T_m$), then a constant temperature $T_b$ ($T_b<T_m$) is imposed on the bottom wall ($x=0$) to drive the freezing process. 

According to the work of Alexiades et al. \cite{AlexiadesBook2018}, the analytical solution of this problem can be given by,
\begin{equation}
	T(x, t)=\left\{\begin{array}{ll}
		T_b-\left(T_b-T_{\mathrm{m}}\right) \frac{\operatorname{erf}\left(x / 2 \sqrt{\alpha_s} t\right)}{\operatorname{erf}(k)} & 0<x<X_i(t), t>0 \\
		T_0 & x>X_i(t), t>0
	\end{array},\right.
\end{equation}
where $\operatorname{erf}(x)=\frac{2}{\sqrt{\pi}} \int_0^x e^{-\eta^2} d \eta$ is the error function, $\operatorname{erfc}(x)=1-\operatorname{erf}(x)$ is the complementary error function, $X_i(t)=2 k \sqrt{\alpha_s t}$ is the location of the solid-liquid interface, the parameter $k$ is the root of the following transcendental equation,
\begin{equation}
	\frac{C_{p, s}\left(T_m-T_{b}\right)}{L \exp \left(k^2\right) \operatorname{erf}(k)}=k \sqrt{\pi} .
\end{equation}
For this problem [see in Fig. \ref{fig2}(a)], the Dirichlet boundary condition is applied in the $x$ direction, and some physical parameters are set as $T_b=-1, T_m=0, T_0=0, C_{p,l}/C_{p,s}=1, L=10$. We conduct some simulations, and present the results in Fig. \ref{fig2}(b) and Fig. \ref{fig2}(c), in which the  comparisons of the temperature distributions and the evolutions of the solid-liquid phase-change interface between the numerical results and analytical solutions under different solid-liquid density ratios are shown. It can be clearly seen that the temperature gradually increases from the bottom boundary to the freezing interface, while remaining a constant in the liquid phase. It is also found that the numerical results are in good agreement with analytical solutions, which demonstrates the accuracy of the preset LB method.

\begin{figure}[H]
	\centering
	\includegraphics[scale=0.5]{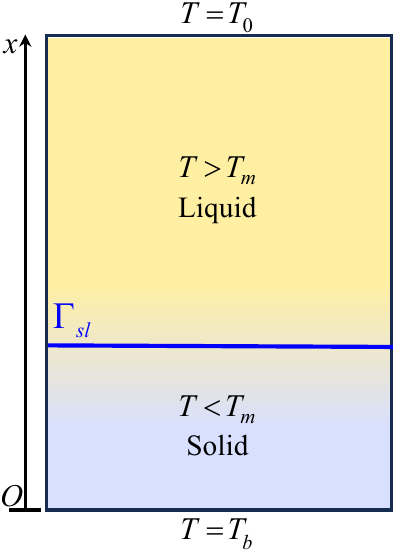} 
	\label{fig3a}
	\put(-110 ,120){(\textit{a})}
	\quad
	\includegraphics[scale=0.85]{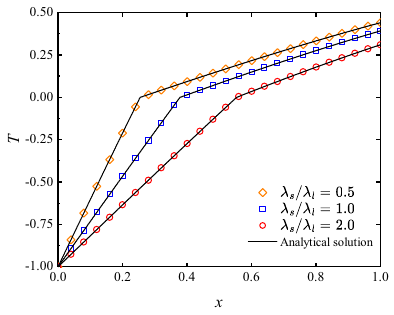} 
	\label{fig3b}
	\put(-165,120){(\textit{b})}
	\includegraphics[scale=0.85]{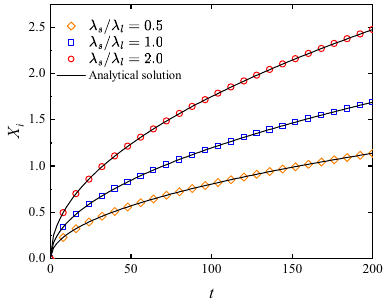} 
	\label{fig3c}
	\put(-160,120){(\textit{c})}
	\caption{ Schematic diagram of the two-phase freezing by the conduction (a). Comparisons of the temperature distribution $T$ (b) and the solid-liquid interface evolution (c) between the numerical results and the analytical solutions under different values of thermal conductivity ratio $\lambda_s/ \lambda_l$. }
	\label{fig3}
\end{figure}

As a general case, the two-phase freezing by conduction is also considered. As shown in Fig. \ref{fig3} (a), the entire domain is initially filled with liquid phase and keeps at the initial temperature $T_0$ ($T_0>T_m$), and the temperature of bottom wall is fixed to be $T_b$ ($T_b<T_m$) when $t>0$. Then the analytical solution of the temperature can be expressed as \cite{AlexiadesBook2018} 
\begin{equation}
	T(x, t)= \begin{cases}T_b-\left(T_b-T_m\right) \frac{\operatorname{erf}\left(x / 2 \sqrt{\alpha_s t}\right)}{\operatorname{erf} k} & 0<x<X_i(t), t>0 \\ T_0+\left(T_m-T_0\right) \frac{\operatorname{erfc}\left(x / 2 \sqrt{\alpha_l t}\right)}{\operatorname{erfc}\left(k \sqrt{\alpha_s / \alpha_l}\right)} & x>X_i(t), t>0\end{cases},
\end{equation}
where $X_i(t)=2 k \sqrt{\alpha_s t}$ is the location of phase interface, the parameter $k$ is the root of the following transcendental equation \cite{AlexiadesBook2018},
\begin{equation}
	\frac{C_{p, s}\left(T_m-T_{b}\right)}{L \exp \left(k^2\right) \operatorname{erf}(k)}-\frac{C_{p, l}\left(T_0-T_m\right) \sqrt{\alpha_l / \alpha_s}}{L \exp \left(k^2 \alpha_s / \alpha_l\right) \operatorname{erfc}\left(k \sqrt{\alpha_s / \alpha_l}\right)}=k \sqrt{\pi} .
\end{equation}
In the following simulations, the physical parameters are given as $T_b=-1, T_m=0, T_0=1, C_{p,l}/C_{p,s}=1, L=10$. For many phase-change materials, the thermal conductivity of the solid phase differs from that of the liquid phase. Therefore, to test the capacity of the present LB method in handling phase-change materials with different thermal properties in the solid and liquid phases, we perform some numerical simulations, and show the results in Figs. \ref{fig3}(b) and \ref{fig3}(c). From these two figures, one can see that under different thermal conductivities, the numerical results of temperature distribution at $t=10$ and the evolution of the solid-liquid interface agree well with the analytical solutions.

\subsection{Three-phase Stefan problem}

\begin{figure}[H]
	\centering
	\includegraphics[scale=0.51]{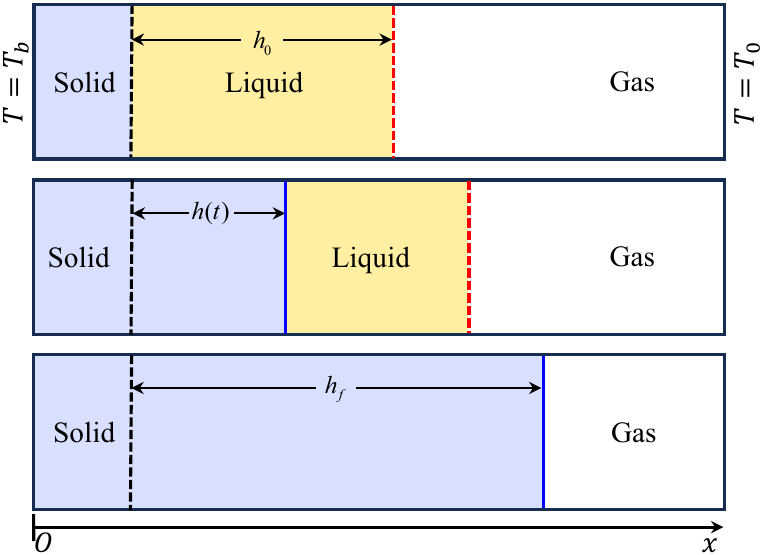} 
	\put(-195 ,130){(\textit{a})}
	\quad
	\includegraphics[scale=0.93]{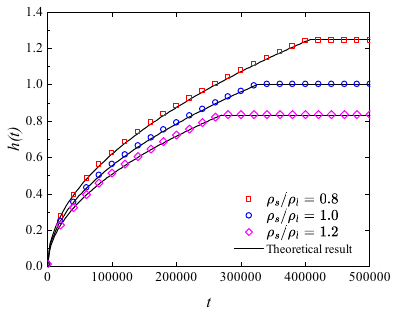} 
	\put(-180,130){(\textit{b})}
	\caption{ The schematic of the three-phase Stefan problem (a), from left to right, shows the initial stage before freezing, the intermediate stage where part of the liquid phase has been frozen into the solid phase, and the final moment when the freezing is complete. The evolutions of the freezing front position $h(t)$ under different values of solid-liquid density ratio $\rho_s/\rho_l$ (b). }
	\label{fig4}
\end{figure}

In the above problems, the influence of the volume change during the freezing process has been neglected, now we test the ability of present LB method in prediction of volume change during the freezing process by simulating the three-phase Stefan problem. The schematic of this problem is depicted in Fig. \ref{fig4}(a), where the domain is initially filled with gas, liquid, and solid phases, the initial height of the liquid phase is denoted as $h_0$. The temperature of the system is initially maintained at $T_0$, and a constant temperature $T_b$ ($T_b<T_m$) is imposed on the bottom wall at time $t>0$, causing the liquid phase to begin freezing. After the liquid phase is completely frozen, the height of the solidified liquid is equal to $h_f$. In our simulations, the initial profiles of the order parameters are given by
\begin{equation}
	\begin{aligned}
		& \phi_0(x, y)=0.5+0.5 \tanh \frac{2\left(h_s-y\right)}{\varepsilon}, \\
		& \phi(x, y)=\left[1-\phi_0(x, y)\right] \tanh \frac{2\left(h_0-y\right)}{\varepsilon},
	\end{aligned}
	\label{eq_initial}
\end{equation}
where $h_s$ and $h_0$ are the initial heights of the solid and liquid phases, respectively. The physical parameters are given by $\bar{M}=0.1, \sigma=0.001, \varepsilon=4, \mu=0.1, T_b=-1$, $T_m=0$, $T_0=1$, $C_{p,s}/C_{p,l}=1$, $\lambda_{s}/\lambda_{l}=1$ and $\Delta x=\Delta t=1$. The periodic boundary condition is applied in $y$ direction, while the no-flux, Dirichlet and no-slip boundary conditions for phase, temperature and flow fields are imposed in $x$ direction. Based on the work of Lyu et al. \cite{LyuJCP2021}, the evolution of the freezing front height can be given as $h_s(t)=2 \kappa \sqrt{\alpha_s t}$, where $\kappa$ is a root of the following transcendental equation,
\begin{equation}
	\frac{Ste }{\exp \left(\kappa^2\right) \operatorname{erf}(\kappa)}-\frac{\lambda_l\left(T_0-T_m\right) \sqrt{\alpha_r}  Ste }{\lambda_s\left(T_m-T_w\right) \exp \left(\sqrt{\alpha_r} \kappa \rho_r\right)^2 \operatorname{erfc}\left(\sqrt{\alpha_r} \kappa \rho_r\right)}=\kappa \sqrt{\pi},
\end{equation}
where $ \rho_r=\rho_s / \rho_l$ and $\alpha_r=\alpha_s / \alpha_l$ are the ratios of the density and thermal diffusivity between the solid and liquid phases, respectively. In addition, the convection velocity induced by volume change is determined by $u_y=\frac{d h_s(t)}{d t}\left(1-\frac{\rho_s}{\rho_l}\right)$.

We first present a comparison of the numerical results and theoretical solutions of the freezing front under different values of the solid-liquid density ratio $\rho_s/ \rho_l$ in Fig. \ref{fig4}(b) where $Ste=0.1$. From this figure, one can observe that the numerical results are in good agreement with the theoretical solutions. Specifically, for the case $\rho_s/\rho_l<1.0$, the volumetric expansion occurs, while for the case of $\rho_s/\rho_l>1.0$, the volumetric shrinkage happens, which are consistent with the theoretical prediction. In addition, based on the mass conservation, the final solid height $h_f$ can be determined by the initial liquid height $h_0$ and the solid-liquid density ratio $\rho_s/\rho_l$, i.e., $h_f=\rho_l h_0 / \rho_s $ \cite{ShetabivashJCP2020}. We also conduct a comparison between the numerical data and theoretical solution of solid height $h_f$ in Table \ref{table1}, where the maximum relative error is less than $1.5\%$. These results indicate that the present LB method can accurately capture volume change during the freezing process while maintaining the mass conservation.

\begin{table}[H]
	\centering
	\caption{ A comparison of the final solid height $h_f$ (normalized by the initial height of liquid) between the numerical and analytical data. }	
	\begin{tabular}{ccccccccc}
		\hline \hline
		\multirow{2}{*}{$\rho_s/\rho_l$} & \multirow{2}{*}{Analytical} & \multicolumn{3}{c}{Numerical}                           &  & \multicolumn{3}{c}{Relative error}                      \\ \cline{3-5} \cline{7-9} 
		&                             & $Ste=0.1$ & $Ste=0.15$ & $Ste=0.2$ &  & $Ste=0.1$ & $Ste=0.15$ & $Ste=0.2$ \\ \hline
		0.6                              & 1.6667                      & 1.6551           & 1.6574            & 1.6567           &  & $0.6955\%$       & $0.5552\%$        & $0.5980\%$       \\
		0.7                              & 1.4286                      & 1.4249           & 1.4246            & 1.4243           &  & $0.2580\%$       & $0.2797\%$        & $0.2992\%$       \\
		0.8                              & 1.2500                      & 1.2478           & 1.2477            & 1.2476           &  & $0.1726\%$       & $0.1864\%$        & $0.1904\%$       \\
		0.9                              & 1.1111                      & 1.1108           & 1.1108            & 1.1108           &  & $0.0307\%$       & $0.0303\%$        & $0.0314\%$       \\
		1.0                              & 1.0000                      & 1.0010           & 1.0010            & 1.0012           &  & $0.1010\%$       & $0.1031\%$        & $0.1228\%$       \\
		1.1                              & 0.9091                      & 0.9131           & 0.9129            & 0.9120           &  & $0.4398\%$       & $0.4239\%$        & $0.3177\%$       \\
		1.2                              & 0.8333                      & 0.8379           & 0.8388            & 0.8371           &  & $0.5456\%$       & $0.6566\%$        & $0.4547\%$       \\
		1.3                              & 0.7692                      & 0.7757           & 0.7730            & 0.7723           &  & $0.8430\%$       & $0.4917\%$        & $0.4011\%$       \\
		1.4                              & 0.7143                      & 0.7189           & 0.7184            & 0.7190           &  & $0.6459\%$       & $0.5702\%$        & $0.6561\%$       \\
		1.5                              & 0.6667                      & 0.6725           & 0.6736            & 0.6734           &  & $0.8793\%$       & $1.0414\%$        & $1.0065\%$       \\
		1.6                              & 0.6250                      & 0.6284           & 0.6284            & 0.6284           &  & $0.5450\%$       & $0.5387\%$        & $0.5387\%$       \\ \hline \hline
	\end{tabular}
	\label{table1}
\end{table}

\subsection{Droplet freezing on a flat substrate}

It should be noted that the effect of wettability has not been involved in the above tests, in this part we intend to test the capability of the LB method in the study of the freezing dynamics under different contact angles. The droplet freezing on a cold substrate is a fundamental problem that is widely encountered in many engineering applications. It has been shown that the water droplet can eventually freeze into an ice drop with a pointy tip due to the solid-liquid density difference \cite{MarinPRL2014,SchetnikovAJP2015,LyuJCP2021}. In order to demonstrate that the present LB method can also handle such a complex problem, we will perform a number of simulations and compare the numerical results with the available experimental and numerical data \cite{GuoIJTS2024}. In the experimental study \cite{GuoIJTS2024}, a $12 \ \mu \mathrm{L}$ water droplet and a $10 \ \mu \mathrm{L}$ hexadecane droplet are placed on solid walls at the temperatures of $-10 \ { }^{\circ} \mathrm{C}$ and $15 \ { }^{\circ} \mathrm{C}$, respectively, and the droplets start to freeze when they contact with the substrate. According to the physical parameters of water and hexadecane \cite{GuoIJTS2024} (see Table \ref{tab1}), the values of several dimensionless parameters are determined as $Ste=0.13$, $\rho_s/\rho_l=0.92$, and $\theta=87^{\circ}$. In the following simulations, the droplet is initially placed on the substrate, and once the droplet reaches the specified contact angle $\theta$, a low temperature $T_w$ is imposed on the substrate.

\begin{table}[H]
	\centering
	\caption{ Physical properties of water and hexadecane.}	
	\begin{tabular}{lccccc}
		\hline \hline
		\multicolumn{1}{c}{Property}                                                 & \multicolumn{2}{c}{Water} &  & \multicolumn{2}{c}{Hexadecane} \\ \cline{2-3} \cline{5-6} 
		\multicolumn{1}{c}{}                                                         & Liquid       & Solid      &  & Liquid         & Solid         \\ \hline
		Density, $\rho$ $(\mathrm{kg}\ \mathrm{m}^{-3})$                               & 999          & 917        &  & 774            & 883           \\
		Thermal conductivity, $\lambda$ $(\mathrm{W} \mathrm{m}^{-1} \mathrm{~K}^{-1})$ & 0.581        & 2.16       &  & 0.15           & 0.15          \\
		Specific heat capacity, $C_p$ $(\mathrm{J}\ \mathrm{kg}^{-1} \mathrm{~K}^{-1})$ & 4220         & 2100       &  & 2310           & 1800          \\
		Viscosity, $\mu$ $(\mathrm{Pa}\ \mathrm{s})$                                    & 0.003        & -          &  & 0.003          & -             \\
		Surface tension, $\sigma$ $(\mathrm{N}\ \mathrm{m}^{-1})$                       & 0.076        & -          &  & 0.028          & -             \\
		Latent heat, $L$ $(\mathrm{kJ}\ \mathrm{kg}^{-1})$                              & 333.4        & -          &  & 230            & -             \\
		Solidification temperature, $T_m$ $({ }^{\circ} \mathrm{C})$                    & 0            & -          &  & 18             & -             \\ \hline \hline
	\end{tabular}
	\label{tab1}	
\end{table}

\begin{figure}[H]
	\centering
	\includegraphics[width=0.17\textwidth]{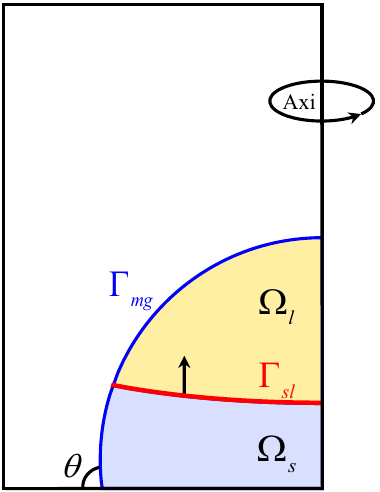} 
	\put(-95 ,92){ (\textit{a})}
	\quad
	\includegraphics[width=0.75\textwidth]{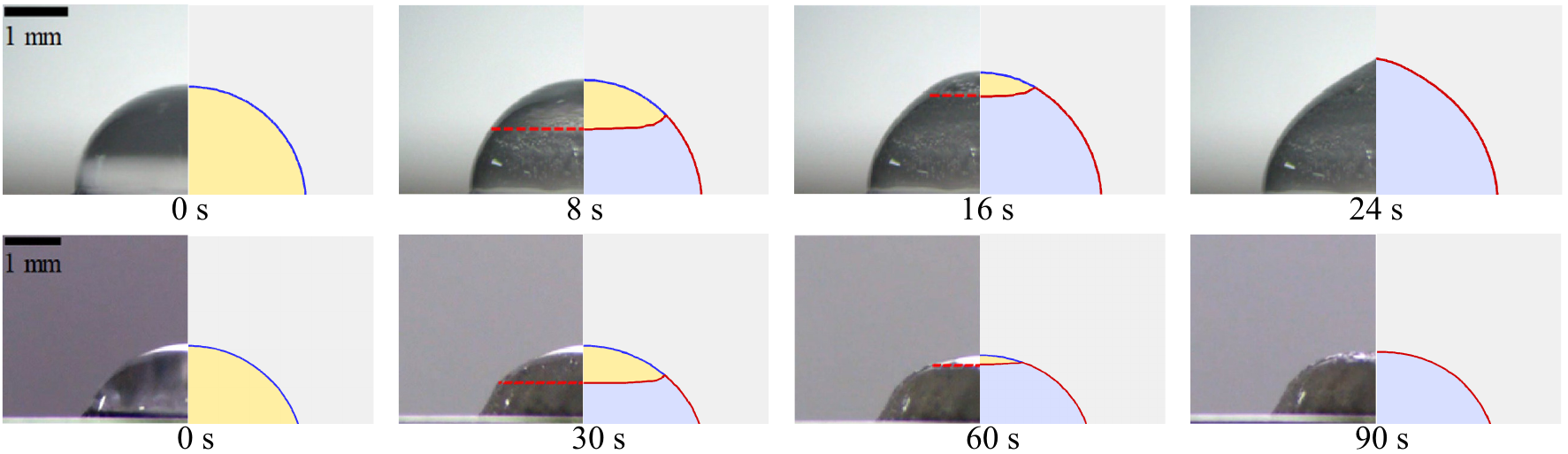} 
	\put(-365,92){(\textit{b})}
	\put(-365,40){(\textit{c})}
	\caption{ Schematic of a freezing water droplet on a supercooled substrate (a). A comparison of experimental \cite{GuoIJTS2024} and numerical results on droplet profiles and solid-liquid phase interfaces of water (b) and hexadecane (c) droplets during the freezing processes. }
	\label{fig5}
\end{figure}

\begin{figure}[H]
	\centering
	\includegraphics[width=0.4\textwidth]{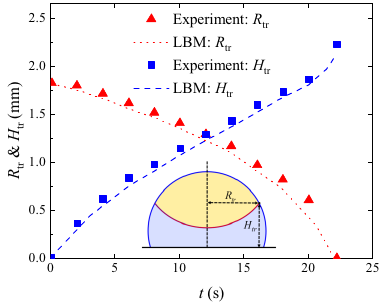} 
	\label{fig4a}
	\put(-190 ,140){(\textit{a})}
	\quad
	\includegraphics[width=0.4\textwidth]{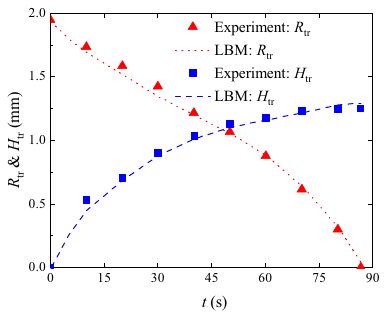} 
	\label{fig4b}
	\put(-190,140){(\textit{b})}
	\caption{ A comparison of the numerical and experimental results \cite{GuoIJTS2024} for the radius $R_{tr}$ and height $H_{tr}$ of three-phase contact line: water (a) and hexadecane (b). }
	\label{fig6}
\end{figure}

We carry out some simulations and present the results in Fig. \ref{fig5}, where a comparison between the numerical results and experimental data for droplet profiles and solid-liquid interfaces at different times is shown. From this figure, it is clear that during freezing process, the freezing front gradually moves from the bottom to the top, forming a concave shape, which is consistent with the previous work \cite{MarinPRL2014}. We note that in the experimental study, the water droplet expands in volume during freezing and eventually forms a conical tip at the top. However, the present numerical results show a smooth and rounded cap instead of a sharp tip, which is consistent with that reported by Lyu et al \cite{LyuJCP2021}. There are two possible reasons for the discrepancy between the numerical and experimental results. The one is that the influence of the contact angle could cause discontinuities in the slope of the liquid-gas-solid triple phase contact line \cite{MarinPRL2014,WangPRL2024}. The other one is that the gas dissolution in the liquid may lead to bubble precipitation and the formation of a porous structure. Actually, Chu et al. \cite{ChuPRF2019} found that there are numerous microbubbles in frozen droplets, supporting the second reason. Under the influence of such a porous structure, the density of ice could decrease to the critical value of $\rho_s/\rho_l = 0.75$ \cite{SnoeijerAJP2012}, thereby forming a sharp ice tip. Recently, Wei et al. \cite{WeiJCP2024} further confirmed this hypothesis through numerical simulations. In their simulations, when the solid-liquid density difference is associated with the concentration of air dissolved in water, the droplet ultimately forms a sharp tip. However, when the solid-liquid density ratio is fixed at $\rho_l/\rho_s = 0.92$, the sharp tip is replaced by a smooth and rounded surface. On the other hand, for the hexadecane droplet, it shrinks inward during freezing due to an increase in density, eventually forming a platform at the top, as reported in some previous numerical studies \cite{HuangPRE2024,MohammadipourJFM2024}. From above discussion, it can be found that the present LB method can capture the freezing characteristics of droplets with different density ratios. In addition, Fig. \ref{fig6} gives a quantitative comparison between numerical results and experimental data in terms of $R_{tr}$ and $H_{tr}$, which represent the radius and height of the three-phase contact line. The results also demonstrate that the proposed LB method can accurately simulate the freezing dynamics of wetting droplets.

\subsection{Droplet freezing on a curved substrate}

\begin{figure}[H]
	\centering
	\includegraphics[scale=0.46]{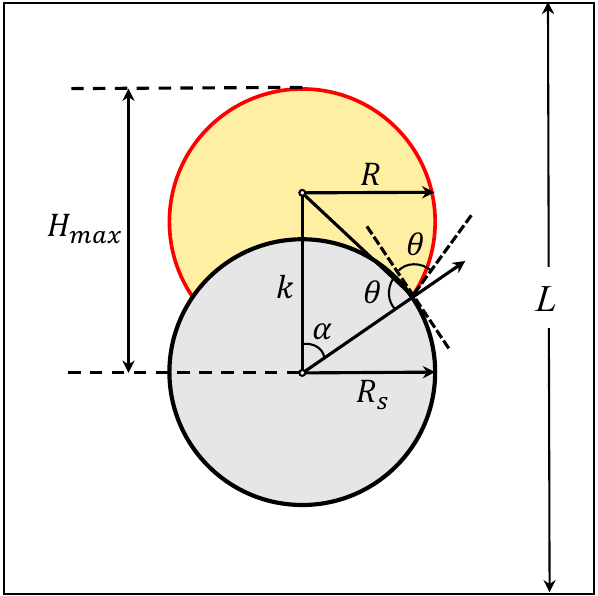}
	\put(-155 ,125){(\textit{a})}
	\quad
	\includegraphics[scale=0.90]{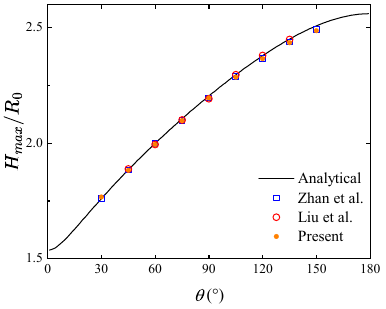}
	\put(-168 ,125){(\textit{b})}
	
	\vspace{0.5cm}		
	\includegraphics[width=0.14\textwidth]{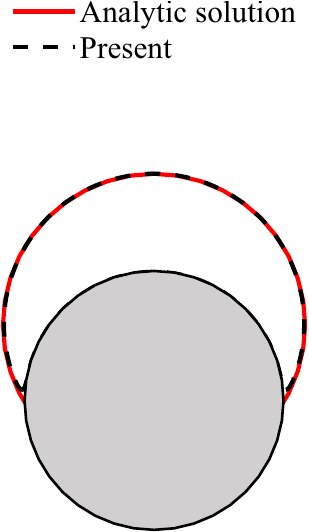} 
	\put(-50,-10){$\theta=30^{\circ}$}
	\put(-80 ,108){(\textit{c})}
	\includegraphics[width=0.14\textwidth]{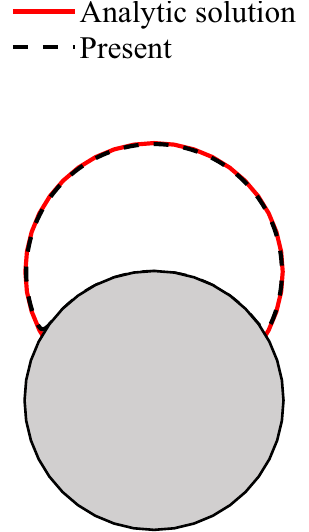}
	\put(-50,-10){$\theta=60^{\circ}$} 
	\includegraphics[width=0.14\textwidth]{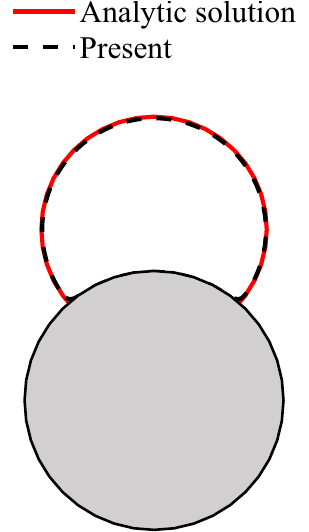} 
	\put(-50,-10){$\theta=90^{\circ}$}		
	\includegraphics[width=0.14\textwidth]{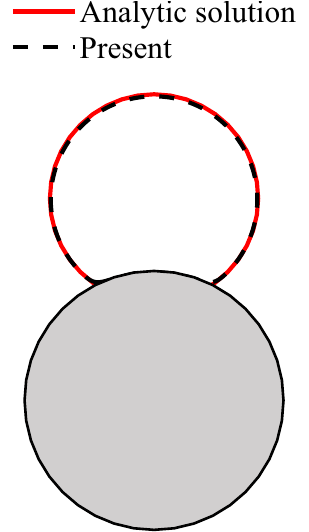} 
	\put(-50,-10){$\theta=120^{\circ}$}		
	\includegraphics[width=0.14\textwidth]{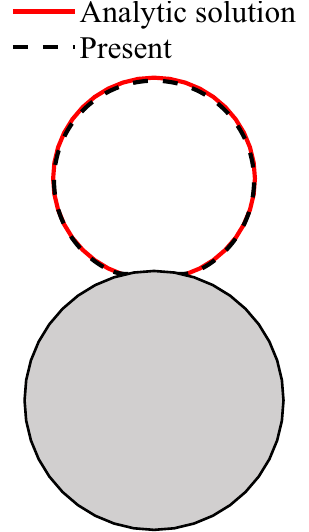} 
	\put(-50,-10){$\theta=150^{\circ}$}
	\caption{ Schematic of a droplet spreading on a cylindrical surface (a). A comparison of the normalized droplet height $H_{max}/R_0$ among the present numerical results, previous data \cite{LiuMMS2022, ZhanJCP2024}, and the analytical solution under different contact angles (b). Predicted equilibrium shape of the droplet spreading on a cylindrical surface. The solid and dashed lines represent results from the present LB method and the analytical solution [Eq. (\ref{analy_contactangle})] under different contact angles. }
	\label{fig7}
\end{figure}

In practical applications, the solid walls are usually curved rather than flat surfaces. As a typical problem, the freezing behavior of sessile droplet on spherical substrate is of particular interest. In this part, we will study the droplet freezing on a cold spherical surface. We first investigate the droplet spreading on a cylindrical surface to test the capacity of the present LB method in predicting the contact angle of droplet on a curved surface, and show the schematic of the problem in Fig. \ref{fig7}(a), where a cylinder with the radius $R_s$ is located at ($L/2,L/4$), and a droplet with radius $R_0$ is initially centered at ($L/2,L/4+R_s$). Here, $\theta$ is the contact angle, $k$ represents the center-to-center distance between the droplet and the cylinder, $H_{max}=k+R$ denotes to the maximum height of the droplet at the equilibrium state. When the droplet is in the equilibrium state, the analytical solution of the order parameter can be expressed as
\begin{equation}
	\phi(x, y)=\frac{1}{2}+\frac{1}{2} \tanh \frac{R-\sqrt{(x-L/2)^2+(y-L/4-k)^2}}{\varepsilon / 2} \text {, }
	\label{analy_contactangle}
\end{equation}
where $k$ is given by $	k=\sqrt{R^2+R_s^2-2 R R_s \cos \theta}.$ Based on the  volume conservation of the droplet, we have
\begin{equation}
	\left(\frac{\pi}{3}+\frac{\sqrt{3}}{2}\right) R_s^2 = (\alpha+\theta) R^2-\alpha R_s^2+R R_s \sin \theta 
\end{equation}
where $\alpha=\cos ^{-1}\left(R_s-R \cos \theta / \sqrt{R^2+R_s^2-2 R R_S \cos \theta}\right)$. According to above formula, one can obtain the values of $k$ and $R$ under different values of contact angle $\theta$.

In the following simulations, the parameters are set as $R_0=R_s=50$, $M=0.1$, $\sigma=0.001$, and $L=300$. Initially, the phase-field variables are given by 
\begin{equation}
	\begin{aligned}
		& \phi_0(x, y)=\frac{1}{2}+\frac{1}{2} \tanh \frac{R_s-\sqrt{(x-L/2)^2+(y-L/4)^2}}{\varepsilon / 2}, \\
		& \phi(x, y)=\left[1-\phi_0(x, y)\right] \tanh \frac{R_0-\sqrt{(x-L/2)^2+\left(y-L/4-R_s\right)^2}}{\varepsilon / 2}.
	\end{aligned}
	\label{initialc}
\end{equation}
The periodic boundary condition is imposed on the left and right boundaries, while the no-flux boundary condition is applied at the top and bottom boundaries. We conduct some simulations and compare the normalized maximum height of the droplet between numerical and analytical solutions in Fig. \ref{fig7}(b). As shown in this figure, the numerical results agree well with the analytical solution and previous data \cite{LiuMMS2022, ZhanJCP2024}. In addition, Fig.  \ref{fig7}(c) presents a comparison of the final shapes of equilibrium droplet, demonstrating that the numerical results are in good agreement with the analytical solutions under different contact angles. These result confirm that the proposed LB method can accurately predict the equilibrium morphology of a droplet on the complex solid surface.

\begin{figure}[H]
	\centering
	\includegraphics[width=0.95\textwidth]{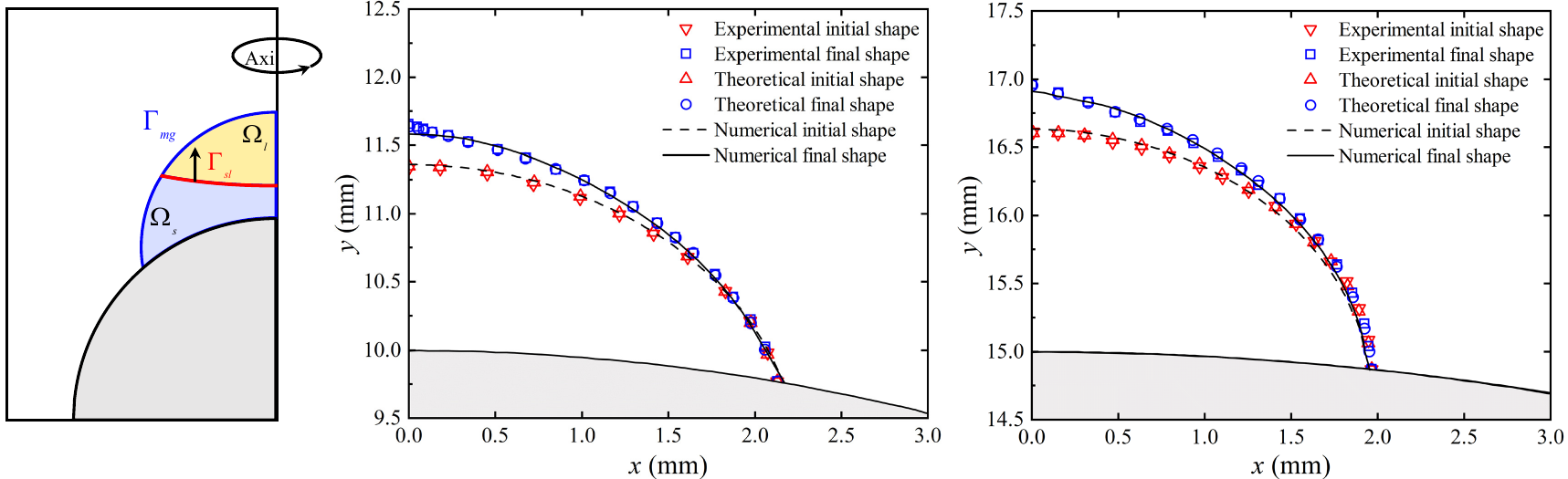} 
	\put(-455,128){(\textit{a})}
	\put(-353,128){(\textit{b})}
	\put(-176,128){(\textit{c})}
	\caption{ Schematic of the droplet freezing on a supercooled cylindrical surface (a). A comparison of initial and final droplet profiles between the theoretical results and experiential data \cite{ZhangATE2024, JuETFS2018} and numerical results at $V_0=12~\mu\mathrm{L}, T_s = -9.5^{\circ}C, \theta=64^{\circ}, D_{\mathrm{s}}=20 \mathrm{~mm}$ (b) and $V_0=13~ \mu \mathrm{L}, T_s = -20.0^{\circ}C, \theta=80^{\circ}, D_{\mathrm{s}}=30 \mathrm{~mm}$ (c).}
	\label{fig8}
\end{figure}

Next, we study the freezing behavior of the droplet on a supercooled spherical surface. In the previous works \cite{ZhangATE2024, JuETFS2018}, a droplet with the volume $V_0$ is deposited onto a supercooled aluminum sphere with the temperature $T_w$, radius $D_s$, and contact angle $\theta$. The experiments where carried out with an ambient temperature of $28.0 \pm 1.0^{\circ} \mathrm{C}$ and a relative humidity of $20 \pm 5 \%$. the surface temperatures are fixed at $T_w=-9.5^{\circ} \mathrm{C}$ and $T_w=-20.0^{\circ} \mathrm{C}$, and the contact angles are $\theta=64^{\circ}$ and $\theta=80^{\circ}$. In addition, the droplet freezing occurred under a condition with no air flow, and the effect of gravity is also neglected. Our simulations are conducted in a two-dimensional domain with $N_x \times N_y=400 \times 400$, a droplet with the radius of $R=40$ initially positioned at the center of a cooled curved substrate, as illustrated in Fig. \ref{fig8}(a). We present the initial and final droplet profiles predicted by the LB method, and give some comparisons with the experimental data in Fig. \ref{fig8} (b) and (c). It is found that the numerical results are consistent with the experimental observation. However, we would also like to point out that similar to the droplet freezing on flat surface, the droplet on the curved surface eventually forms a tip in the experiments, while in the numerical simulations, the sharp tip is replaced by a smooth and rounded cap, which may be caused by the diffuse interface method considered in this work.

\subsection{Freezing in the rough fracture}

\begin{figure}[H]
	\centering
	\includegraphics[width=1.0\textwidth]{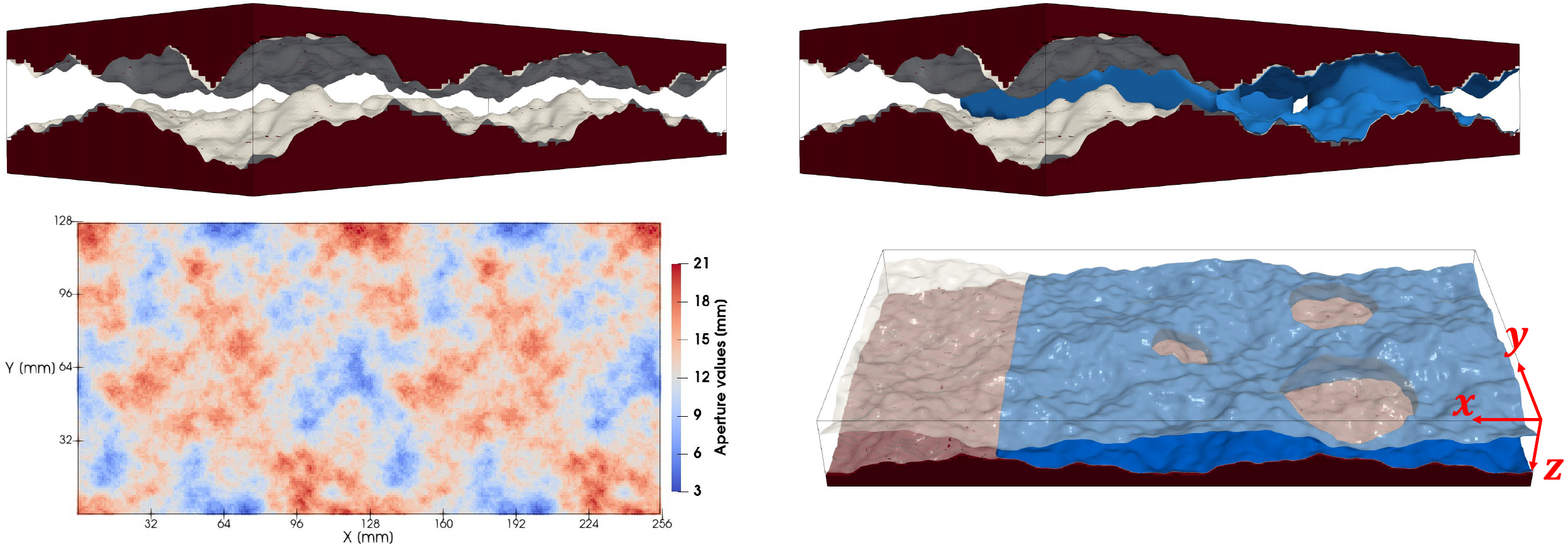} 
	\put(-470,160){(\textit{a})}
	\put(-470,95){(\textit{b})}
	\put(-230,160){(\textit{c})}
	\put(-230,95){(\textit{d})}
	
	\caption{ Examples of fracture generated using pySimFrac with a fractal dimension of 2.25 and an average pore diameter of 12.7 mm. (a). The distribution of pore diameter ranging from 5 to 20 mm (b). The top view (c) and side view (d) of the freezing process within a rough fracture, where the blue part represents the liquid phase and the rest of the region is filled with the gas phase. }
	\label{fig9}
\end{figure}
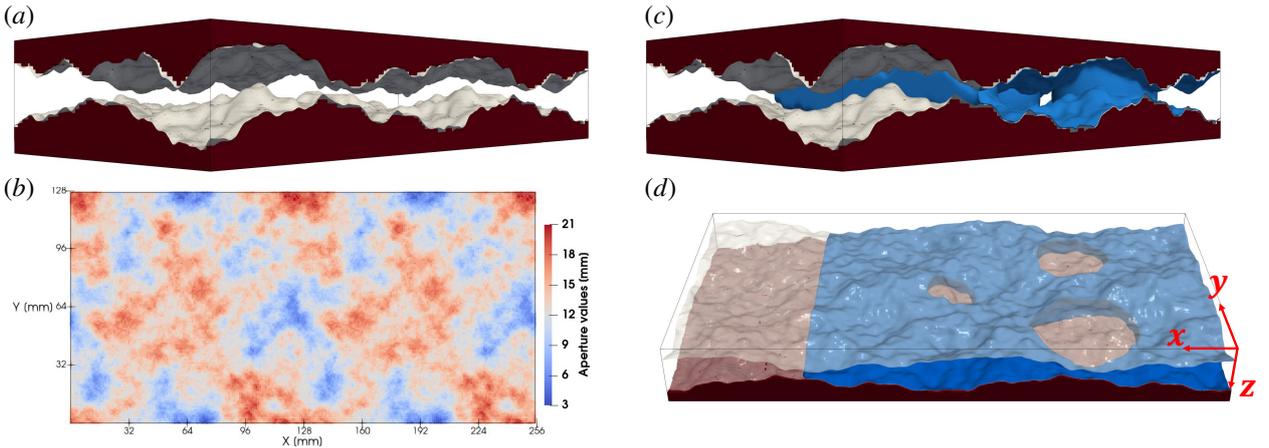

The phenomenon of freezing in the rough fractures, i.e., the fluid freezes within fissures of rocks, has been usually observed in natural processes and human activities. In this part, we consider the process of freezing in a rough fracture to further test the capacity of the present LB method. To generate realistic fracture structure, pySimFrac, a Python toolkit for creating synthetic fractures based on natural analogues, is utilized to numerically generate realistic fractures \cite{GuiltinanWRR2021, GuiltinanCG2024}. Using the Glover method, some fractures with the grid resolution 128 l.u. $\times$ 128 l.u., a standard deviation of 2.5 l.u., a mismatch length of 30 l.u. and different fractal dimensions $F_d$, are created. The geometry is then repeated once in the $x$-direction to obtain a fracture with the grid resolution 256 l.u. $\times$ 128 l.u. $\times$ 30 l.u., Fig. \ref{fig9} (a) and Fig. \ref{fig9} (b) show the volume-rendered fracture image and the corresponding aperture field of the fracture.

In the following simulations, a uniform computational mesh with $Lx \times Ly \times Lz = 256 \ \text{l.u.} \times 128 \ \text{l.u.} \times 30 \ \text{l.u.}$ is applied, and initially, a liquid column with the height $h=0.75Lx$ along the $x$-direction is in the rough fracture, and three circular gas bubbles  with the radii of $0.075Lx$, $0.05Lx$, and $0.01Lx$ are positioned within the liquid column, centered at $(0.25Lx, 0.25Ly, 0.5Lz)$, $(0.5Lx, 0.5Ly, 0.5Lz)$, and $(0.25Lx, 0.75Ly, 0.5Lz)$, respectively. The initial order parameter $\phi$ is given by Eq. (\ref{eq_initial}), the initial distributions of the gas and liquid phases in the fracture are shown in Fig. \ref{fig9}(c) and (d), where the blue part characterizes the liquid phase and the rest part is filled with gas phase. Three gas bubbles are confined, and their interfaces are influenced by the wettability and the fracture geometry. The initial temperature field throughout the entire domain is $T(x,y,z)=T_0$, while the bottom wall is maintained at a low temperature $T(0,y,z)=T_w$, which drives the freezing process. The periodic boundary conditions are applied in the $y$- and $z$- directions, and the bottom and top boundaries in the $x$-direction are treated as two solid surfaces. In addition, the volume expansion caused by density difference usually leads to deformation of the solid structure, but for simplicity, it is not considered here.

\begin{figure}[H]
	\centering
	
	\includegraphics[scale=0.4]{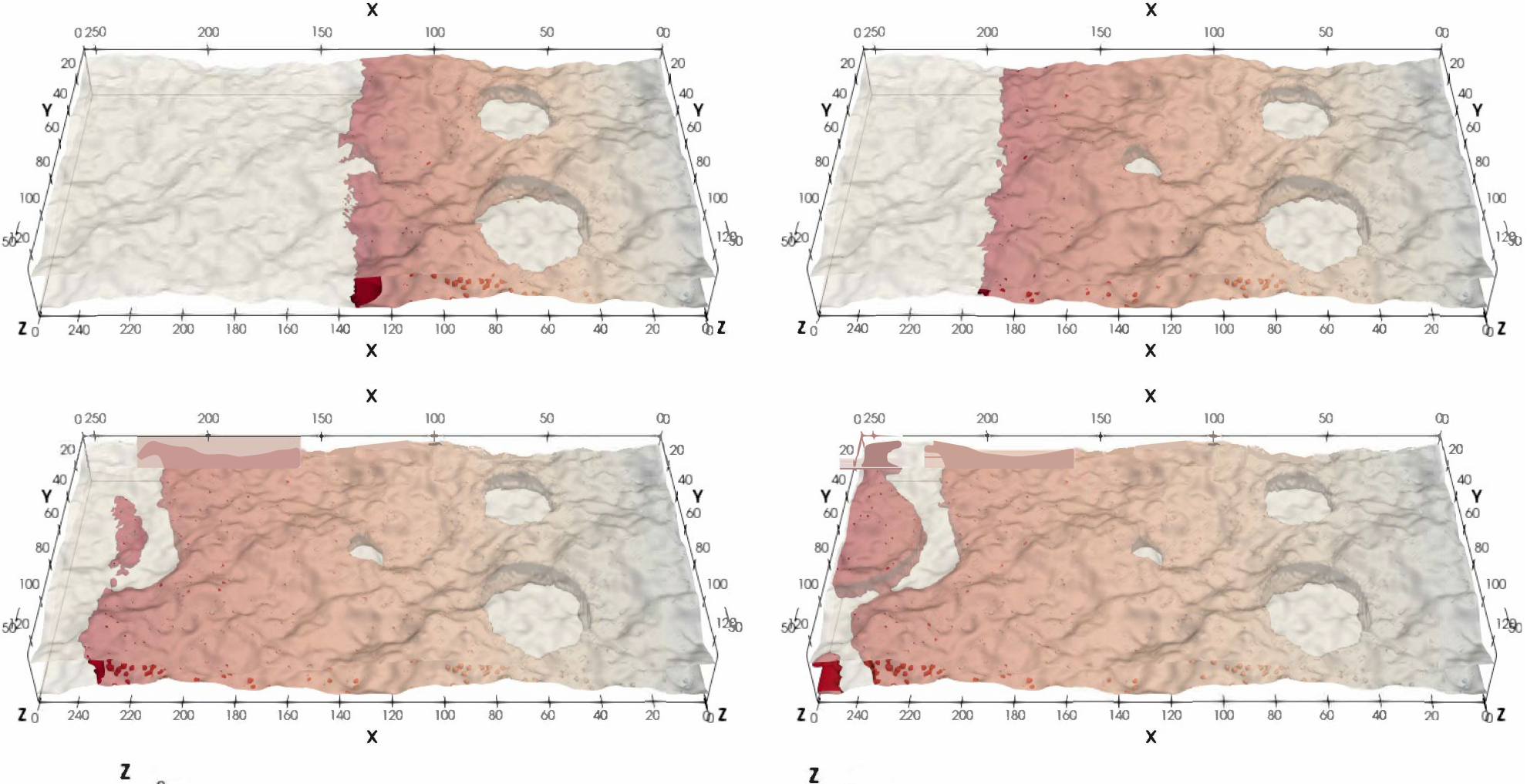} 
	\put(-380 ,180){(\textit{a})}
	\put(-185 ,180){(\textit{b})}	
	\put(-380 ,85){(\textit{c})}
	\put(-185 ,85){(\textit{d})}		
	
	\includegraphics[scale=0.4]{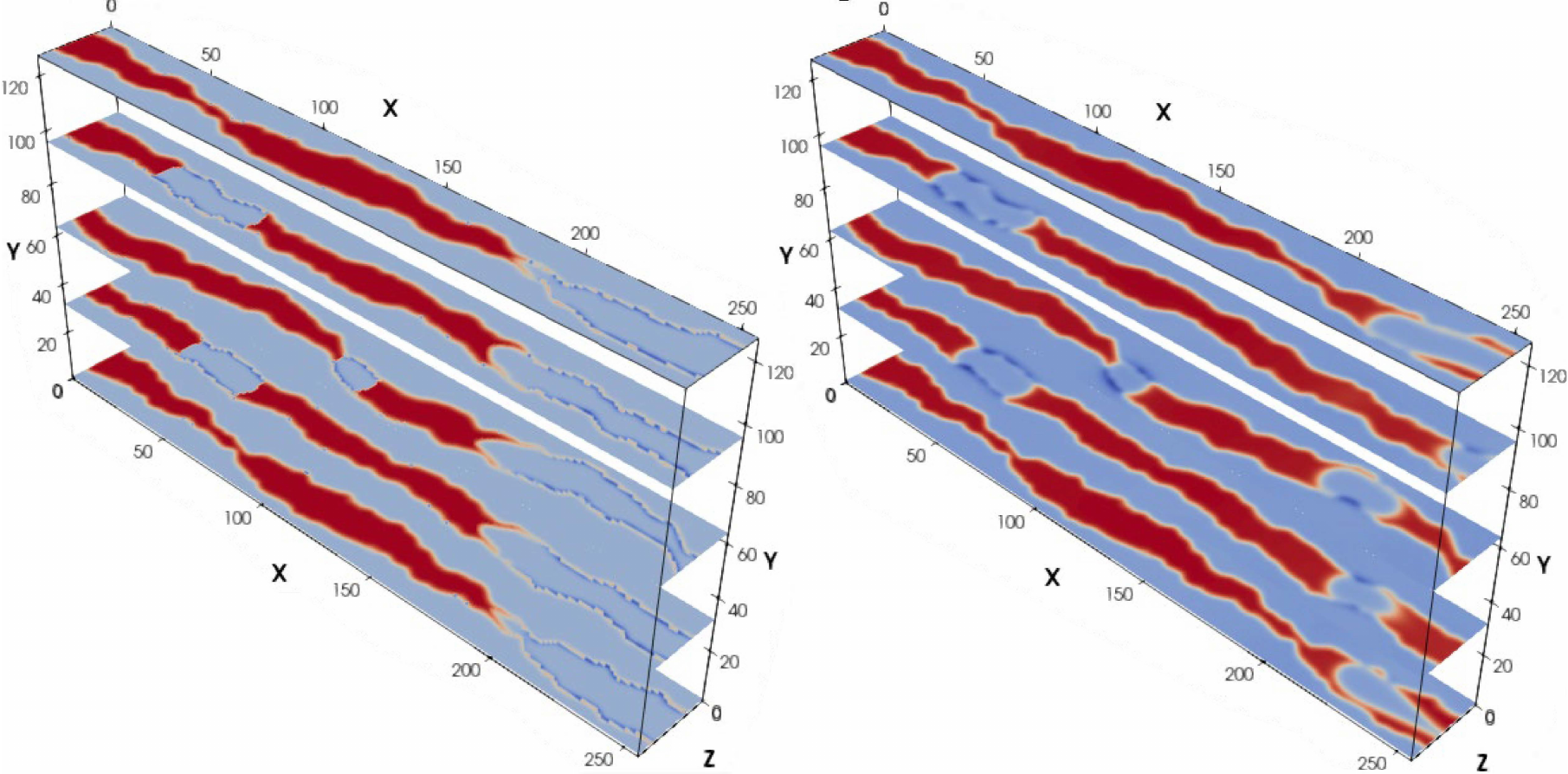} 	
	\put(-365 ,160){(\textit{e})}
	\put(-190 ,160){(\textit{f})}
	
	\caption{ Evolution of the solid fraction $f_s$ during freezing process in a rough fracture at (a) $Fo=0.25$, (b) $Fo=0.5$, (c) $Fo=0.75$ and (d) $Fo=1.0$, the color is used to illustrate temperature distribution. The distributions of solid fractions (e) and order parameter (f) at $Fo=0.5$. }
	\label{fig10}
\end{figure}
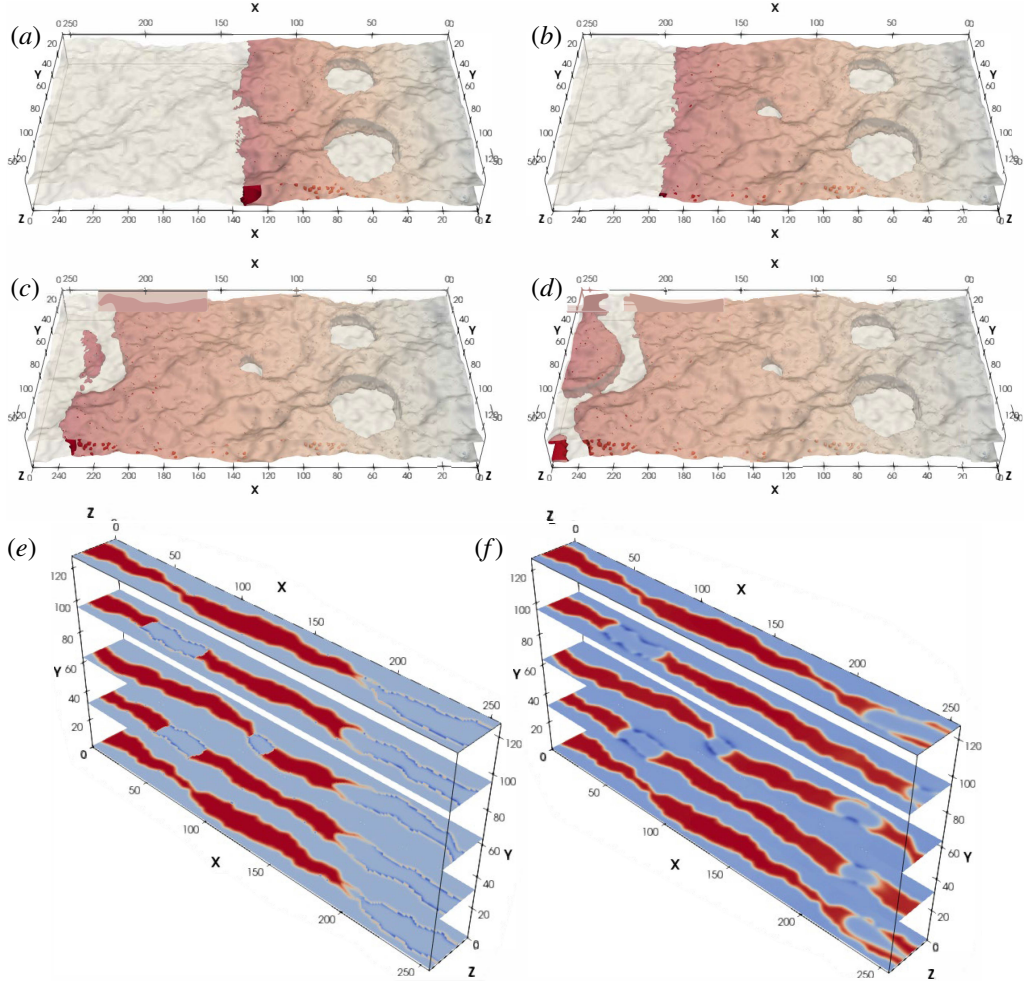

Fig. \ref{fig10} shows the freezing process within the rough fracture, and it is found that the freezing begins at the bottom wall with a low temperature, and propagates upward (along the $x$-direction). Initially, a large temperature gradient accelerates heat transfer and also freezing process. At this stage, the thin ice layer exhibits a low thermal resistance, enabling the rapid heat transfer to the bottom wall and promoting fast freezing. Due to the roughness and heterogeneous wettability of the crack surface, the freezing fronts advance at different rates. The hydrophilic zone, where the liquid is more prone to spread on the solid surface, has a larger thermal conductivity, and liquid phase freezes faster [see Fig. \ref{fig10}(e)], while the hydrophobic zone, where a gas gap may form due to the contraction of liquid film, hinders heat transfer and freezing process. As freezing progresses, the gas bubbles are gradually enveloped by the freezing front, forming isolated gas pockets. The thermal conductivity of gas bubble is much smaller than that of the liquid, which slows down the freezing of the surrounding liquid, creating a localised hysteresis zone. As the thickness of ice layer gradually increases, the temperature gradient decreases, resulting in a slower freezing rate. As a result, the thermal resistance increases, weakening the heat transfer. In addition, under the influence of wettability, the liquid spreads, migrates, or infiltrates along the surface of the crack, forming irregular shapes, as shown in Fig. \ref{fig10}(d) and Fig. \ref{fig10}(f).

\subsection{Freezing in an unsaturated porous medium}
We further consider the freezing process in an unsaturated porous medium to demonstrate the potential of the present LB method for complex freezing problems. We use the real structure of the poorly sorted unconsolidated fluvial sandpack with a porosity $\varphi=0.3585$ \cite{ZhanPD2024}, which features well-separated pores and interconnected structure in the flow direction. To apply the present LB method, the original structure of porous medium needs to be smoothed through the finite-time evolution of a standard CH equation \cite{ZhanPD2024}. In addition, it is also necessary to obtain the distributions of water and gas phases in the unsaturated porous medium before performing numerical simulations. Actually, Pot et al. \cite{PotAWR2015} demonstrated that the LB method can accurately caputure the migration of water, solute, and particle as well as the air-water interfaces in the interstitial pores of porous media, without the need for simplified assumptions about the geometry or topology of soil pores. Following the previous work \cite{PotAWR2015}, we first consider the spontaneous phase separation in the porous medium at the different initial distributions of order parameter $[\phi(x,y)=(1-\phi_0(x,y))\phi_i, \phi_g \leqslant \phi_i \leqslant \phi_l]$ to achieve different values of saturation $S_w$, and then the distribution of the liquid phase can be determined by the balance of cohesive force, adhesive force, and gravity.

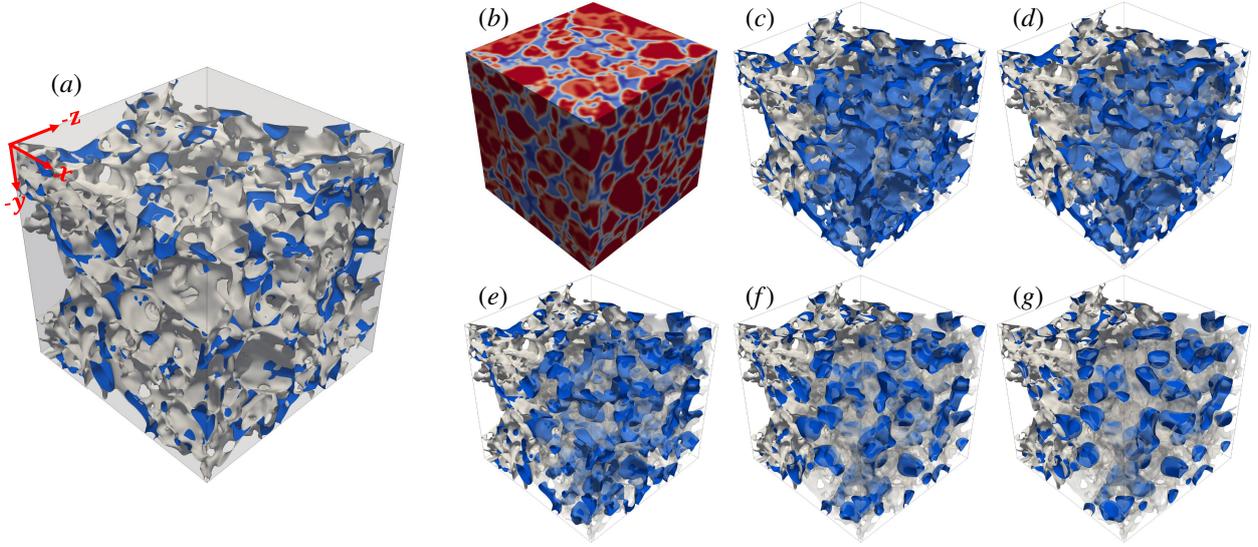
\begin{figure}[]
	\centering
	\includegraphics[width=1.0\textwidth]{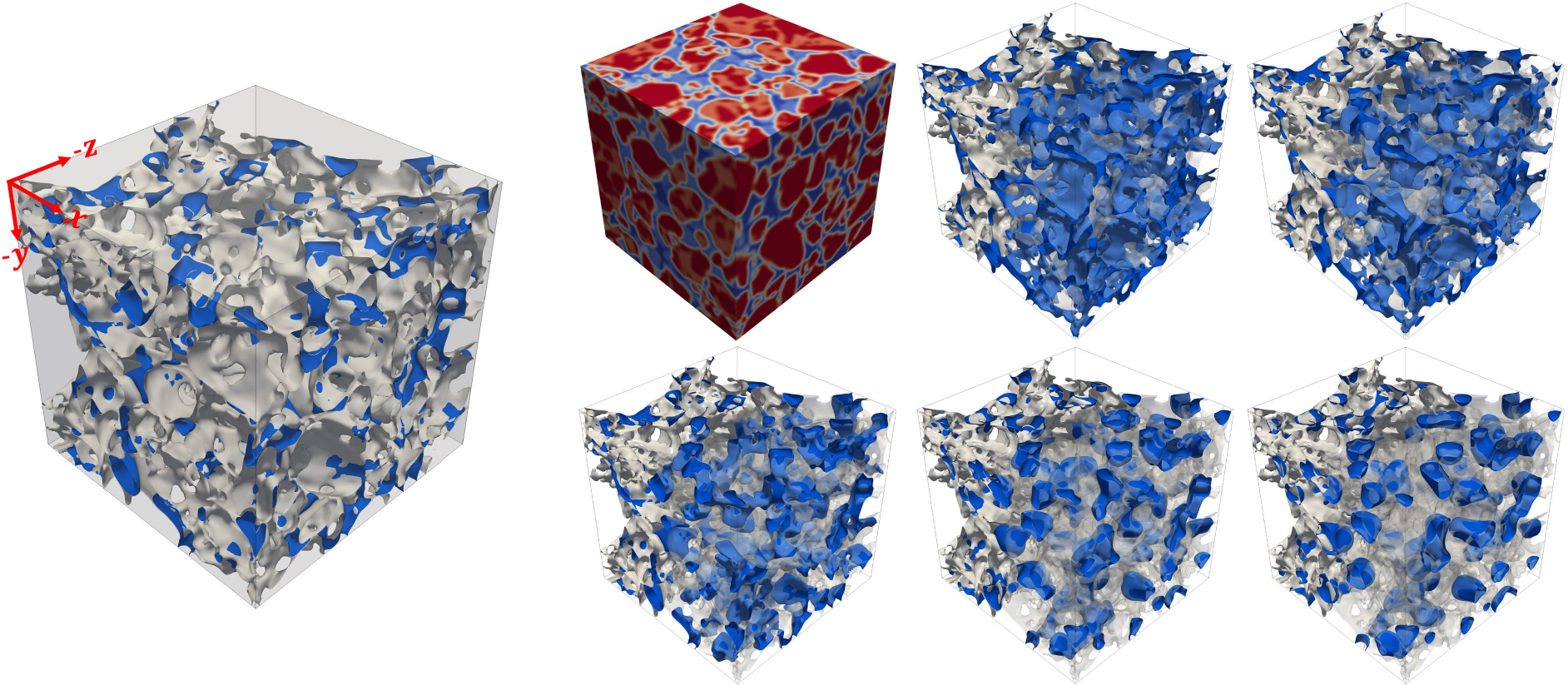} 
	\put(-450,170){(\textit{a})}
	\put(-290,195){(\textit{b})}
	\put(-190,195){(\textit{c})}
	\put(-90,195){(\textit{d})}
	\put(-290,90){(\textit{e})}
	\put(-190,90){(\textit{f})}
	\put(-90,90){(\textit{g})}	
	\caption{ Schematic of the freezing in an unsaturated porous medium (a). The porous medium model used in simulations (b). The equilibrium states of liquid phase under different contact angles: $\theta = 30^{\circ} $ (c), $\theta = 60^{\circ} $ (d), $\theta = 90^{\circ} $ (e), $\theta = 120^{\circ} $ (f), $\theta = 150^{\circ} $ (g) at a water saturation of $S_w = 0.32$. The blue part represents the equilibrium distribution of liquid phase in the porous medium, where skeleton is displayed in semi-transparent gray, and the remaining space is filled with gas phase. }
	\label{fig11}
\end{figure}

Figs. \ref{fig11} (b)-(f) present the equilibrium distributions of the liquid phase under different contact angles $(\theta=30^{\circ},60^{\circ},90^{\circ},120^{\circ},150^{\circ})$, where blue part represents the liquid phase, gray part denotes the skeleton of porous medium, and the remaining space is filled with the gas phase. Under the hydrophilic conditions $(\theta=30^{\circ},60^{\circ})$, the liquid is more prone to spread on the solid surface to form a thin ring (pendular ring), with the liquid–gas interface exhibiting a crescent shape. As the contact angle increases $(\theta=120^{\circ},150^{\circ})$, the pendular rings expand and eventually evolve into isolated liquid droplets. These droplets exhibit nearly spherical liquid–gas interfaces, leading to enhanced the connectivity of gas phase, and the distribution patters of gas and liquid phases under different contact angles are consistent with those reported elsewhere \cite{PotAWR2015,XuWRR2020}. Once the distributions of gas and  liquid phases in the porous medium are obtained, a low temperature $T_b$ is applied to the boundary at $x=0$, the opposite boundary is subjected to a constant temperature $T_w$, and the adiabatic boundary condition is adopted to other boundaries. In this case, the liquid phase starts to freeze from the left to the right side as time goes on. In the following simulations, the physical parameters are kept the same as those stated in the previous problem.

We present the evolution of the solid fraction $f_s$ and temperature distribution in the porous medium in Fig. \ref{fig12} where different contact angles are considered. To see the evolution of freezing front, the colored portion indicates the distribution of the solid fraction, and the porous medium is shown in semitransparent gray. With the increase of time, the solid phase gradually forms in the frozen zone ($T<T_m$), while the liquid and gas phases remain in the non-frozen pores of porous medium ($T>T_m$). As seen from this figure, the freezing front moves from the low-temperature wall to the high-temperature wall, and due to the difference in thermal conductivities of gas and skeleton of porous medium, the development of the freezing front is non-uniform. Specifically, the higher thermal conductivity of skeleton of the porous medium enhances the heat transfer, causing the liquid in contact with the skeleton to reach the freezing temperature much faster, and thereby freezing preferentially. This phenomenon indicates that the freezing process in the unsaturated porous medium is dynamic and heterogeneous, and the thermal conductivity of the skeleton of porous medium has a significant influence on evolution of freezing front.

\begin{figure}[H]
	\centering
	\includegraphics[width=0.45\textwidth]{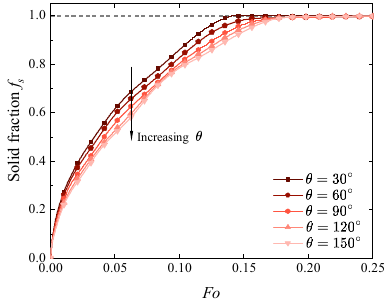}
	\caption{ Evolution of the solid fraction $f_s$ with $Fo$ for different wetting conditions. }
	\label{fig13}
\end{figure}

\begin{figure}[H]
	\centering
	\includegraphics[width=0.22\textwidth]{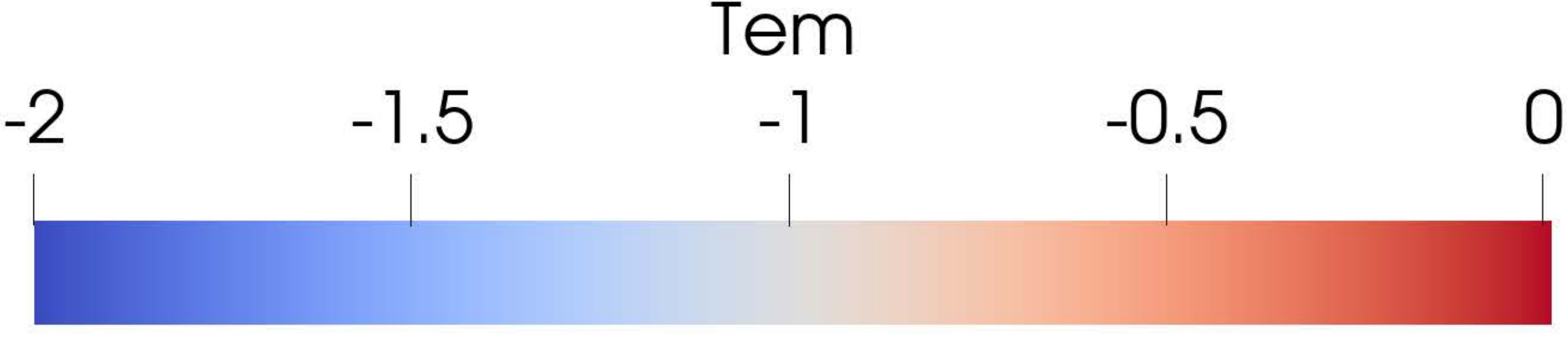} 
	\vspace{0.5cm}	

	\includegraphics[width=0.22\textwidth]{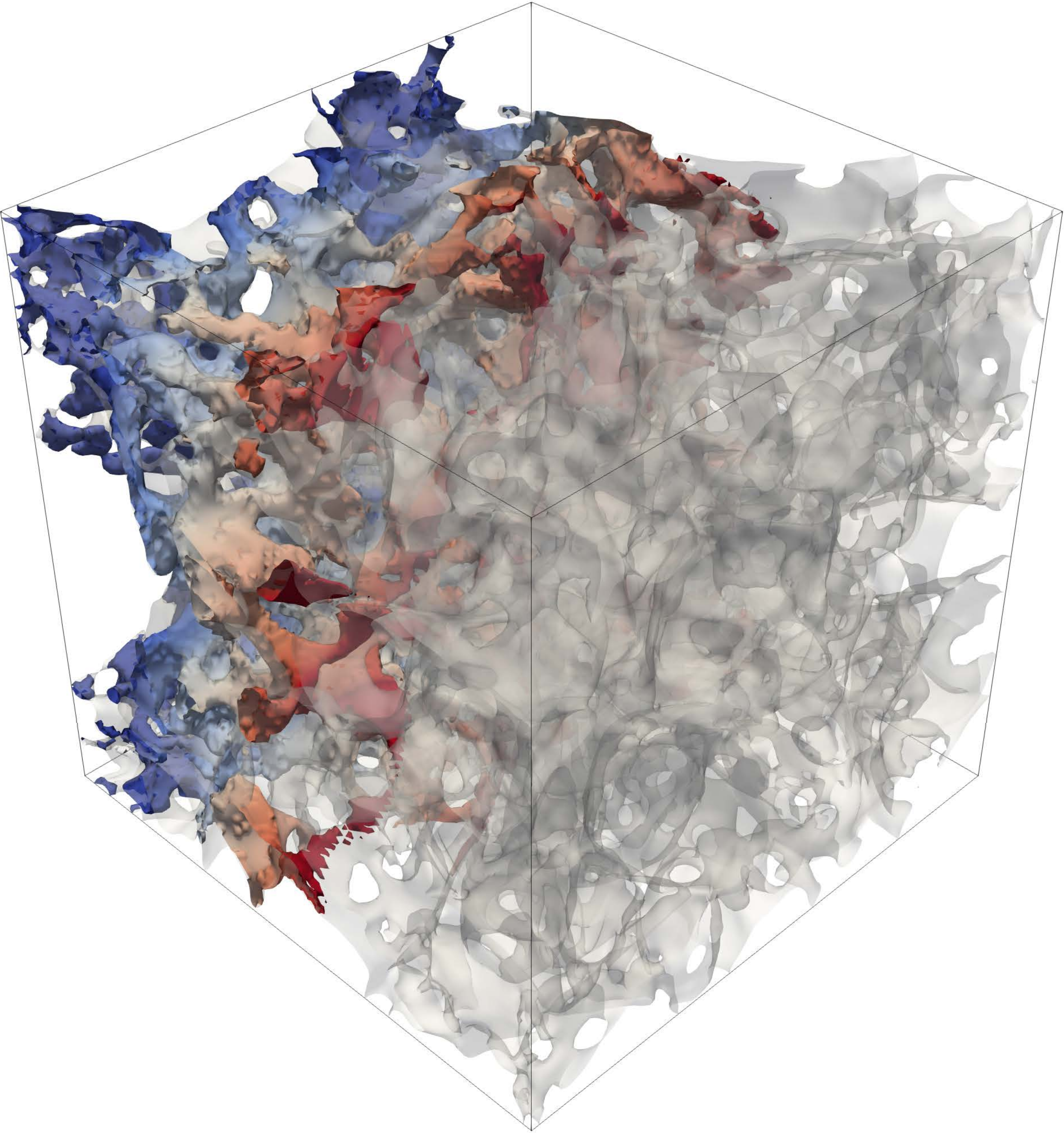} 
	\put(-115 ,60){(\textit{a})}
	\includegraphics[width=0.22\textwidth]{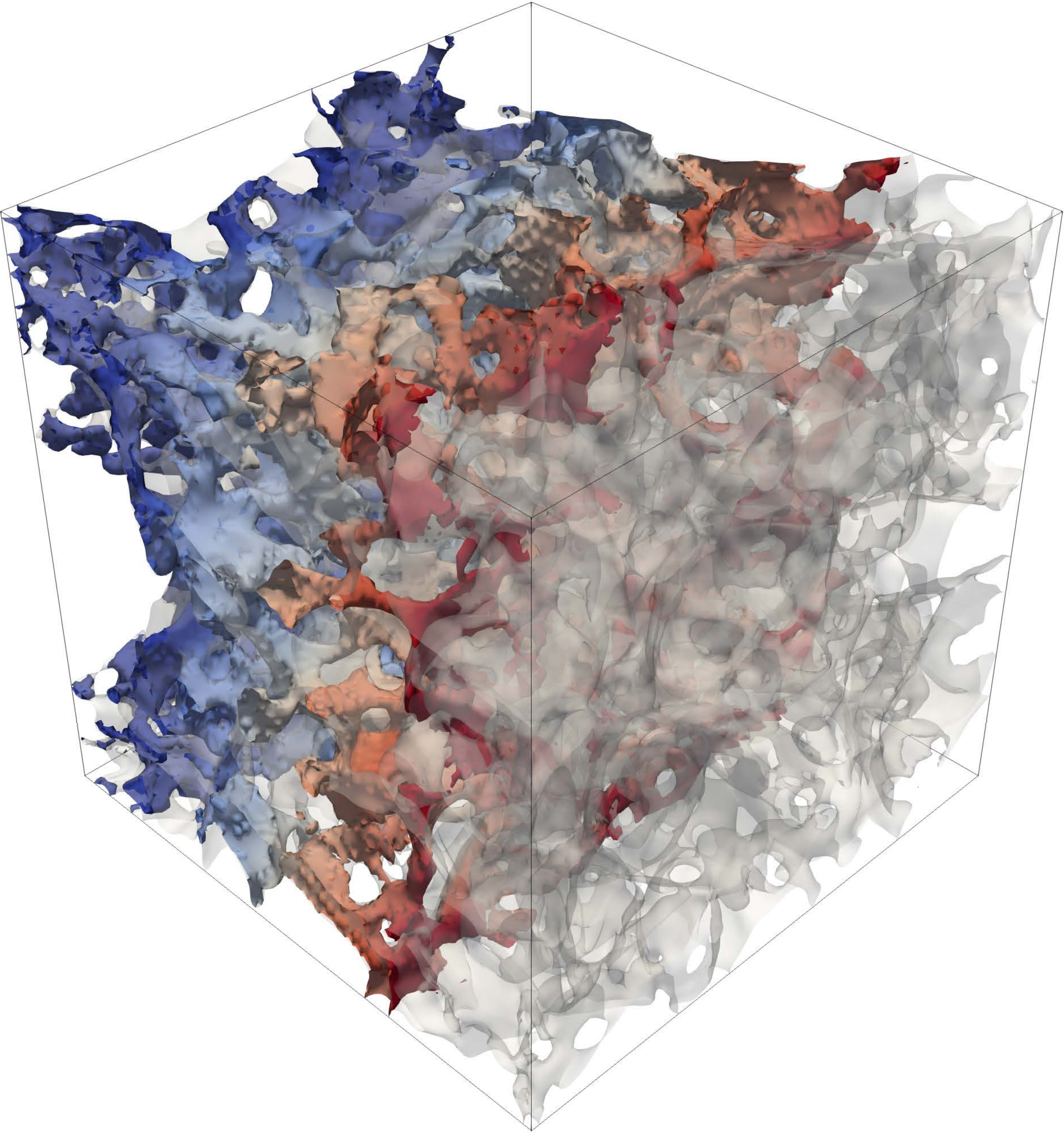}
	\includegraphics[width=0.22\textwidth]{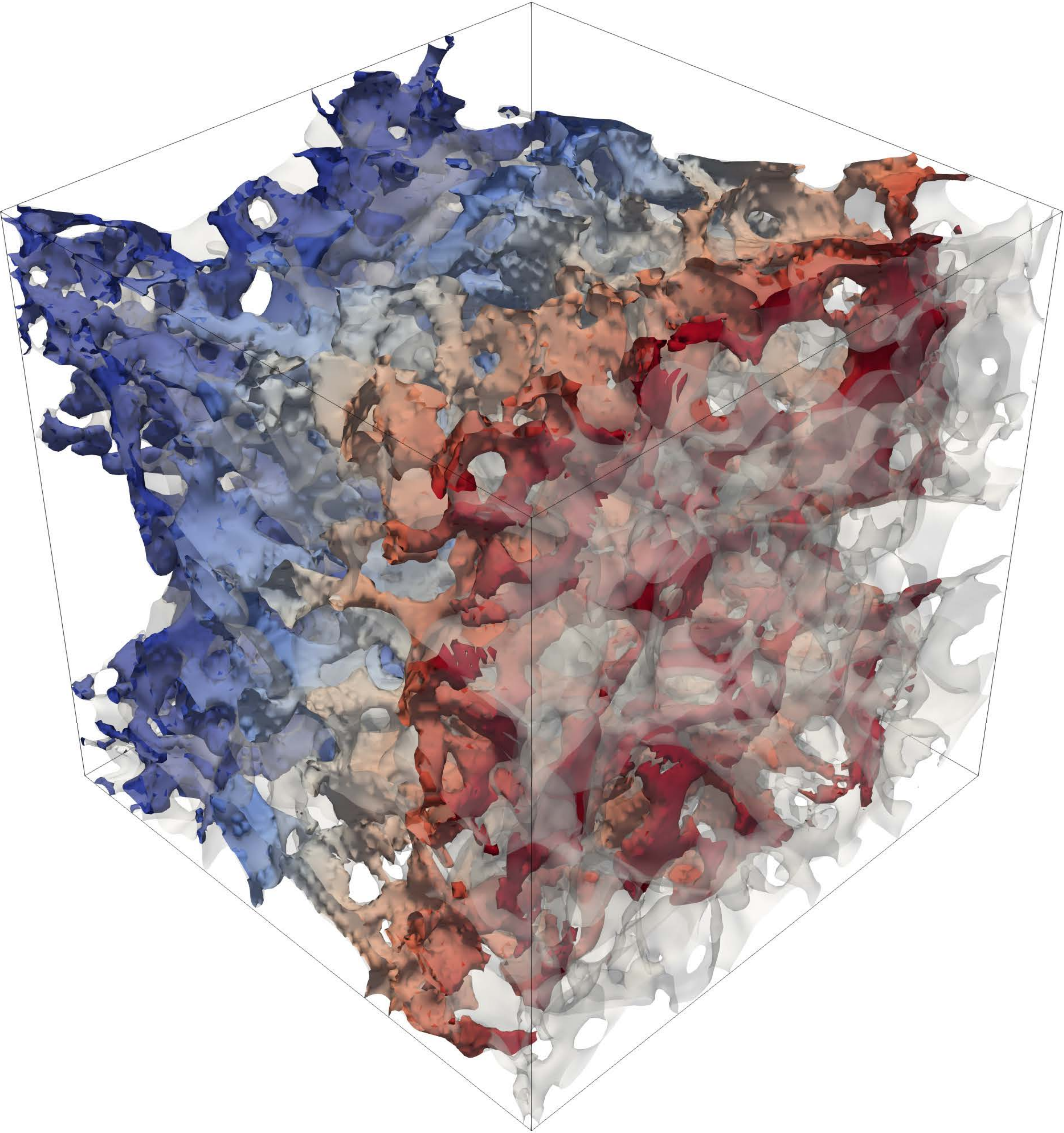} 	
	\includegraphics[width=0.22\textwidth]{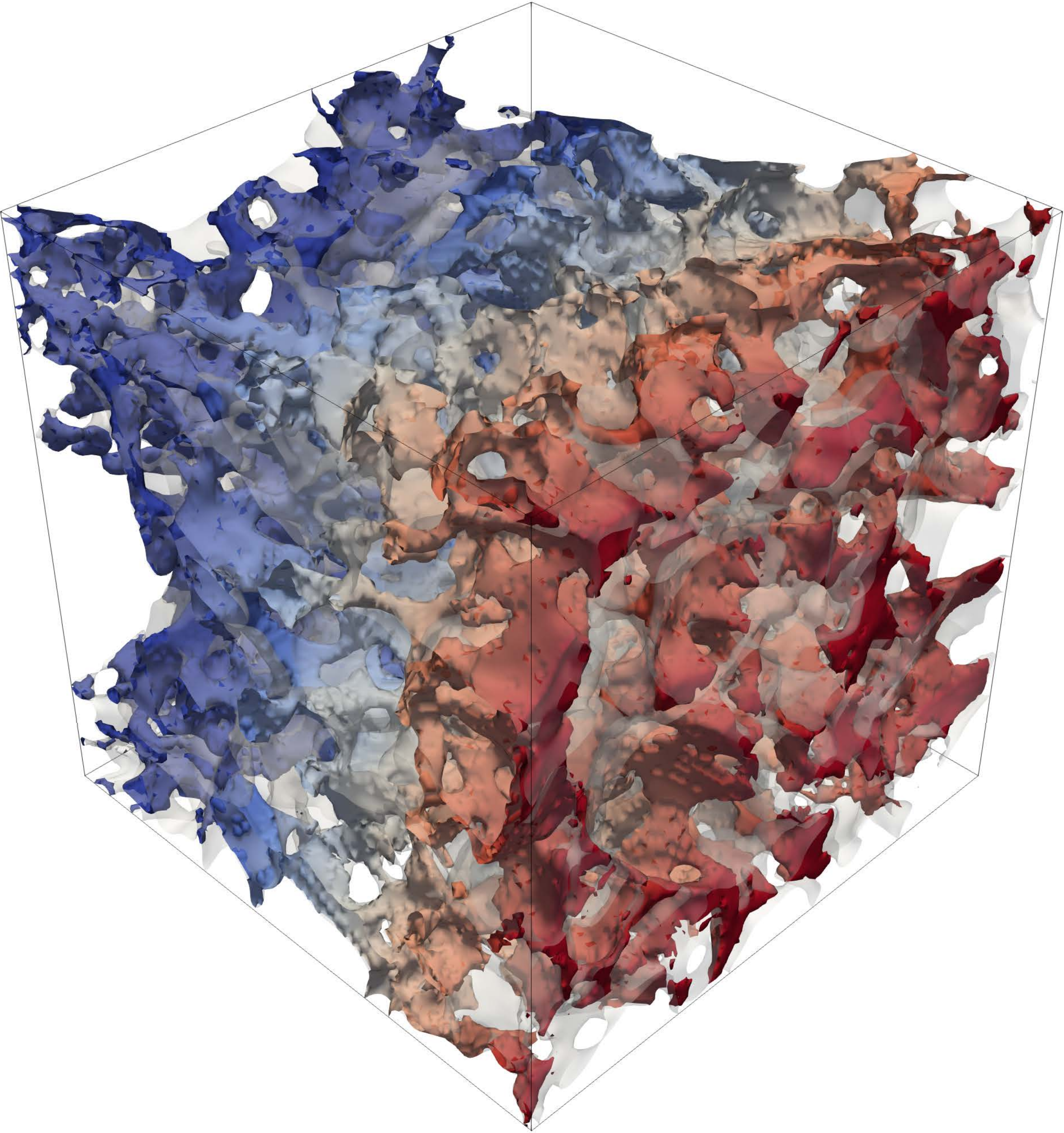}

	\includegraphics[width=0.22\textwidth]{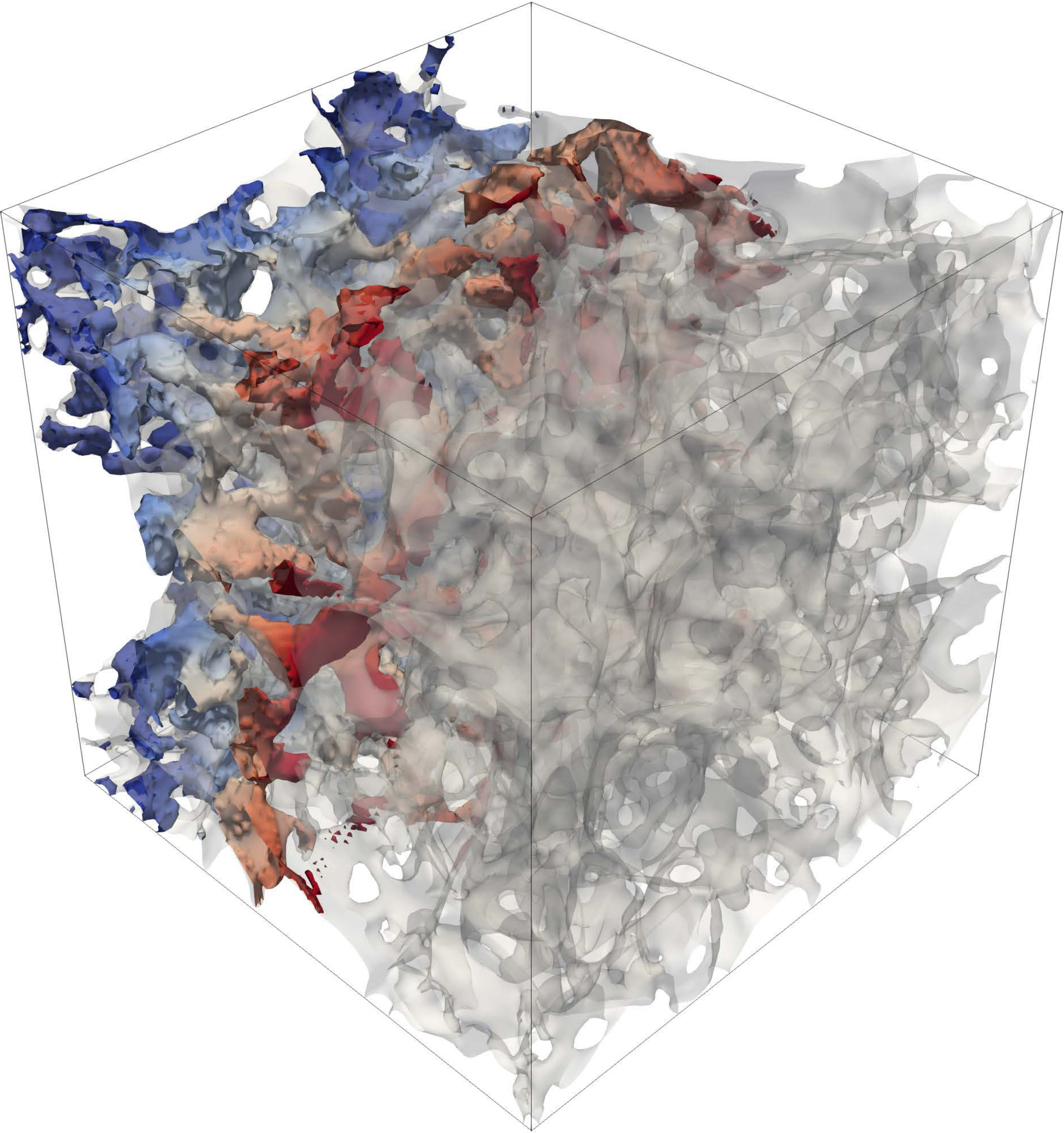} 
	\put(-115 ,60){(\textit{b})}
	\includegraphics[width=0.22\textwidth]{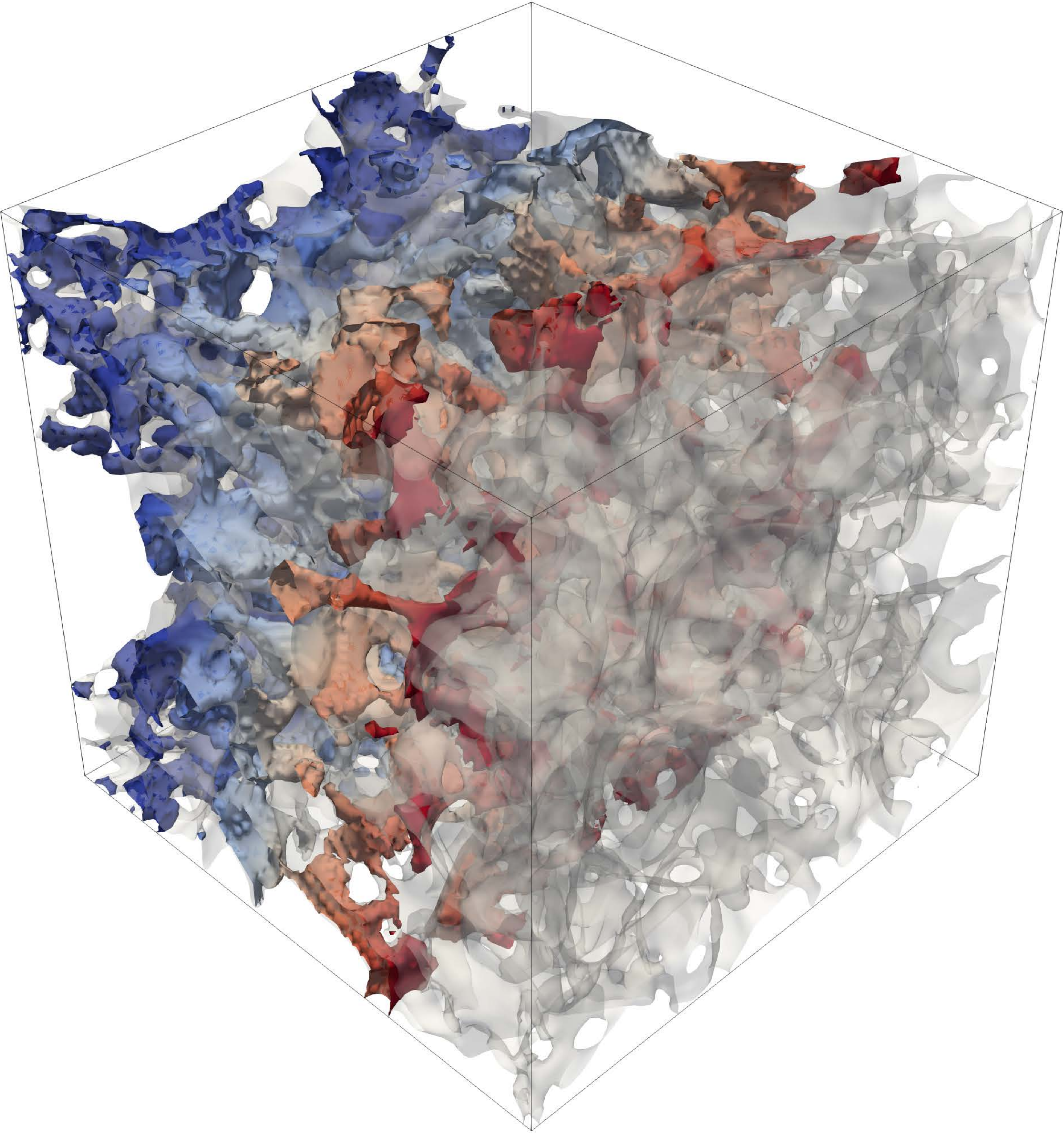}
	\includegraphics[width=0.22\textwidth]{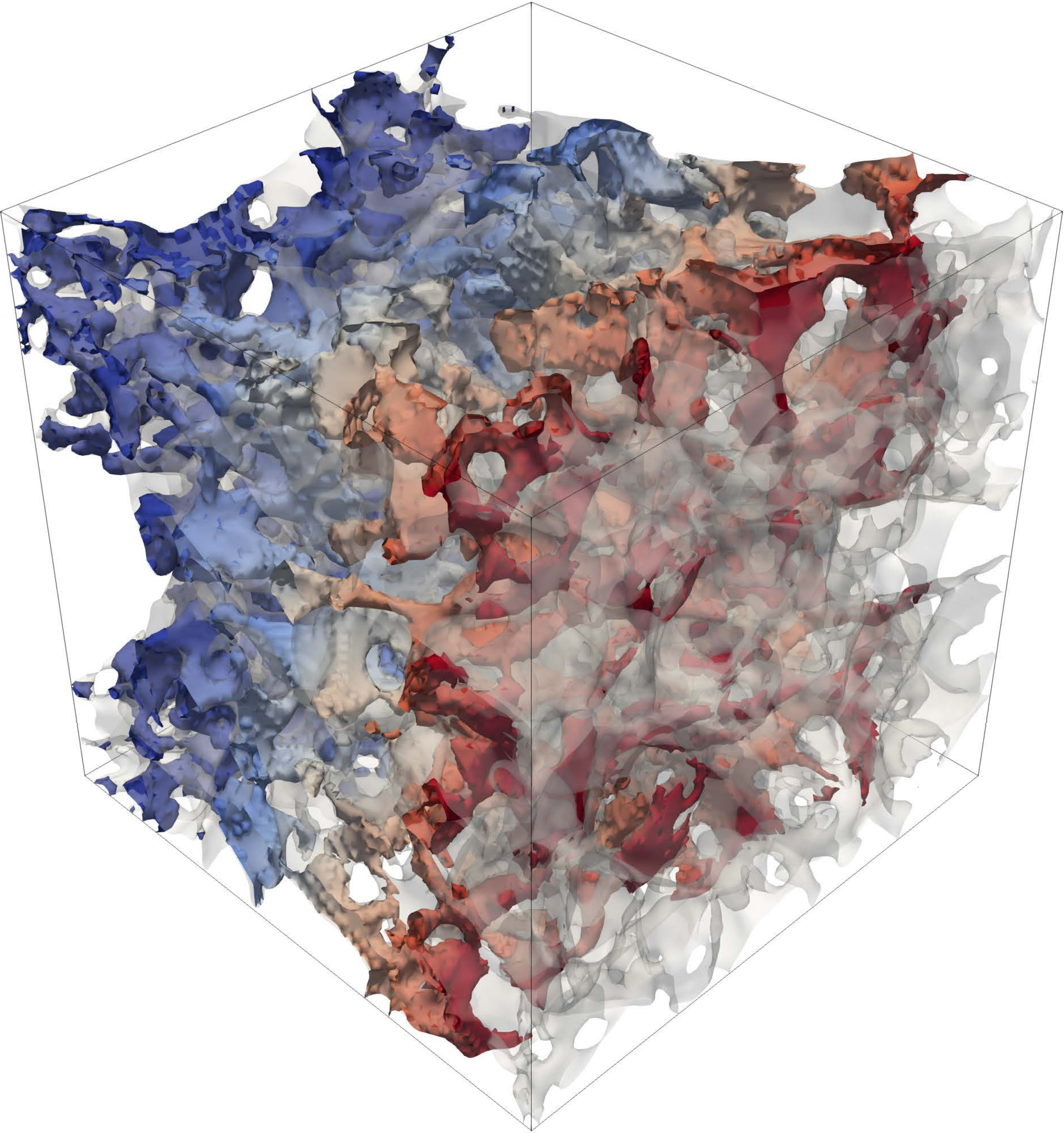} 
	\includegraphics[width=0.22\textwidth]{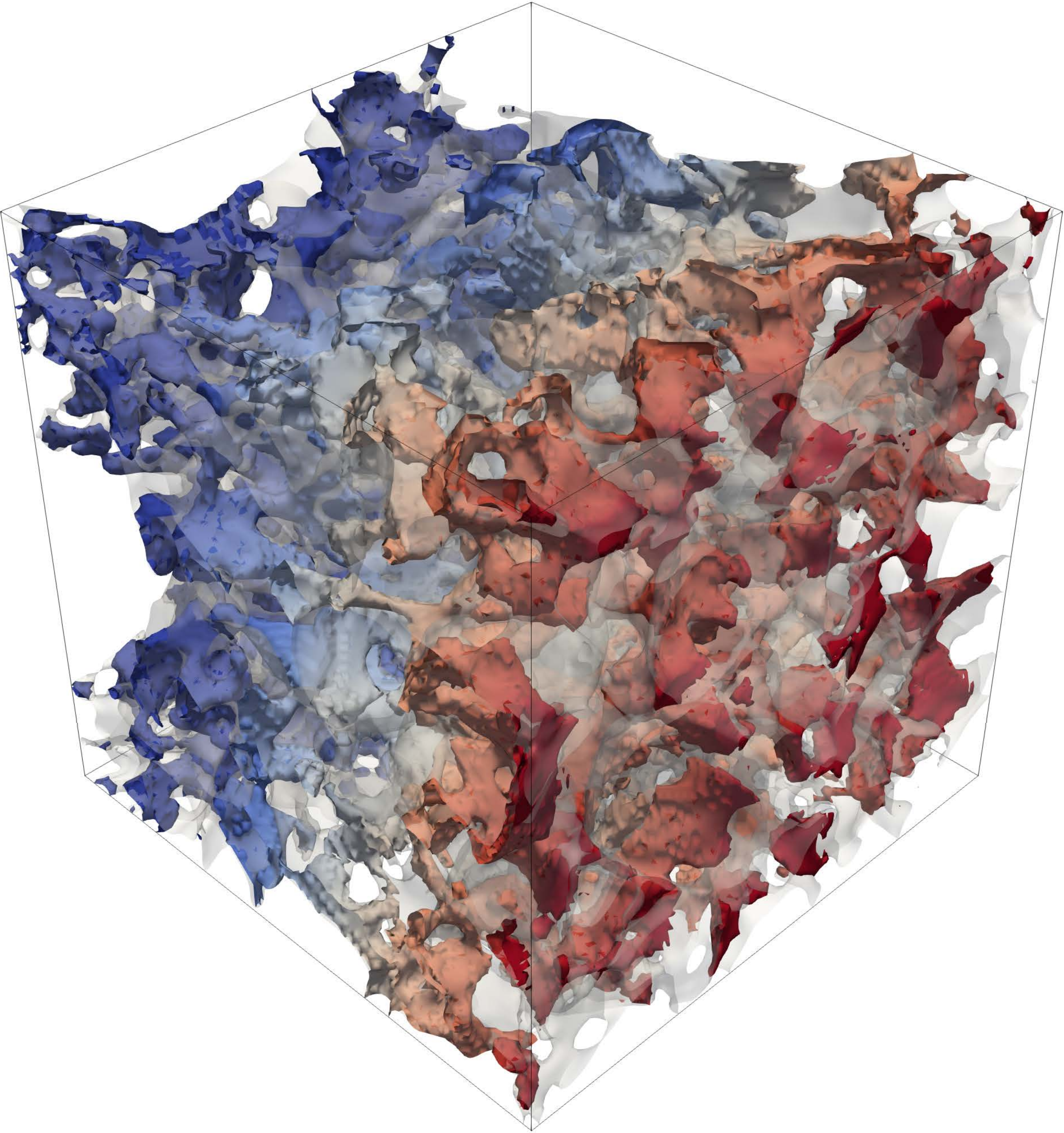}

	\includegraphics[width=0.22\textwidth]{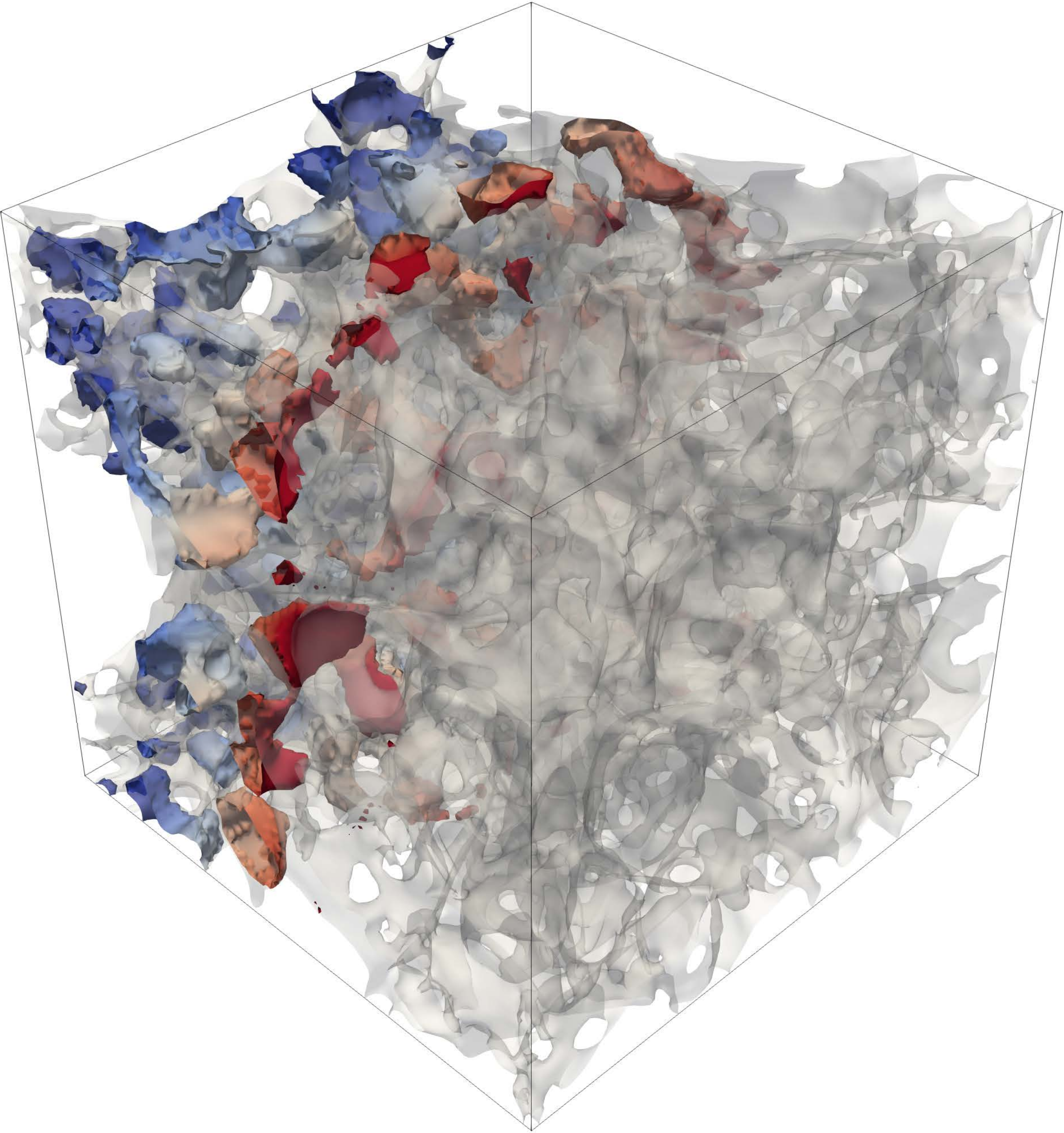} 
	\put(-115 ,60){(\textit{c})}
	\includegraphics[width=0.22\textwidth]{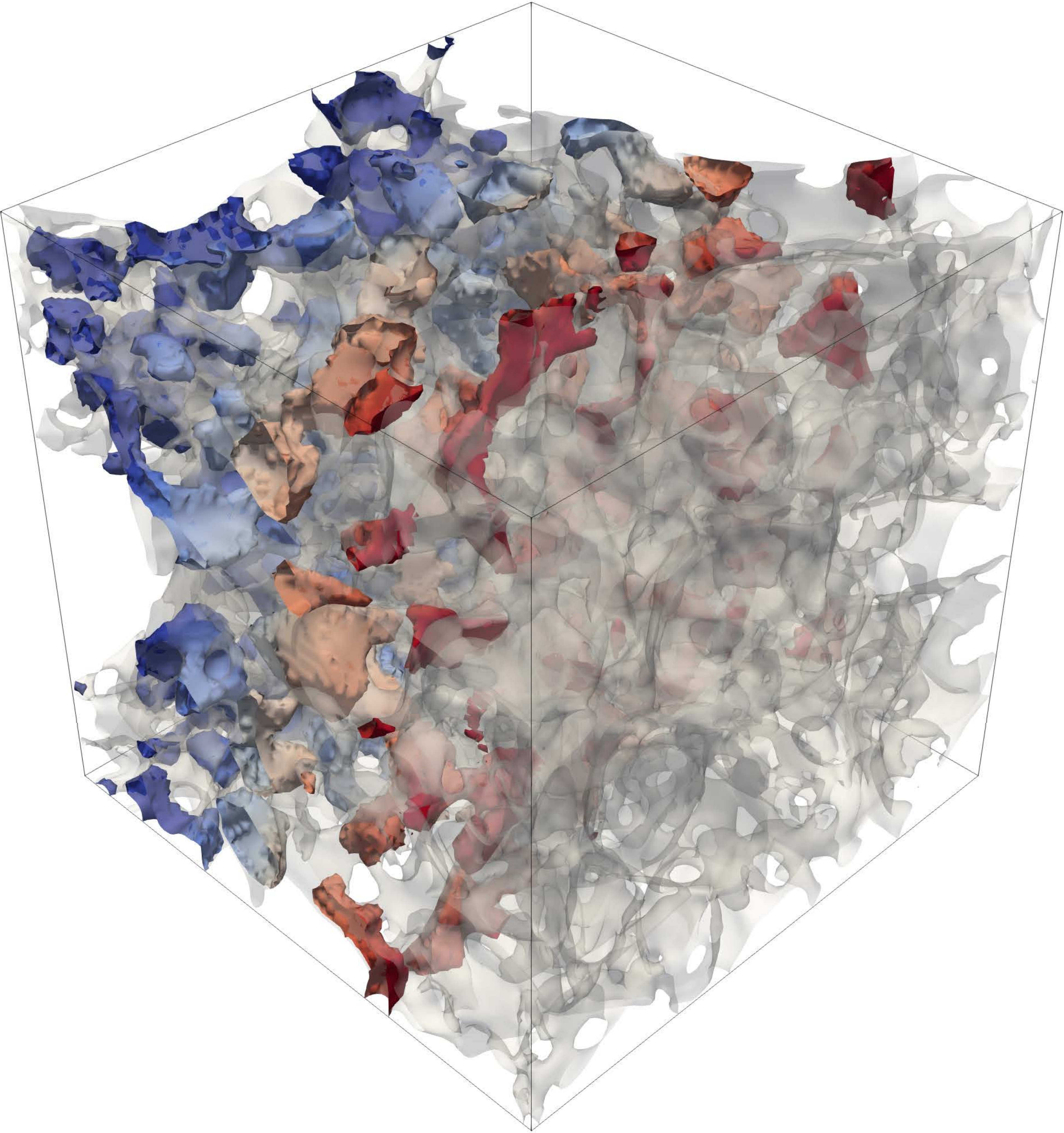}
	\includegraphics[width=0.22\textwidth]{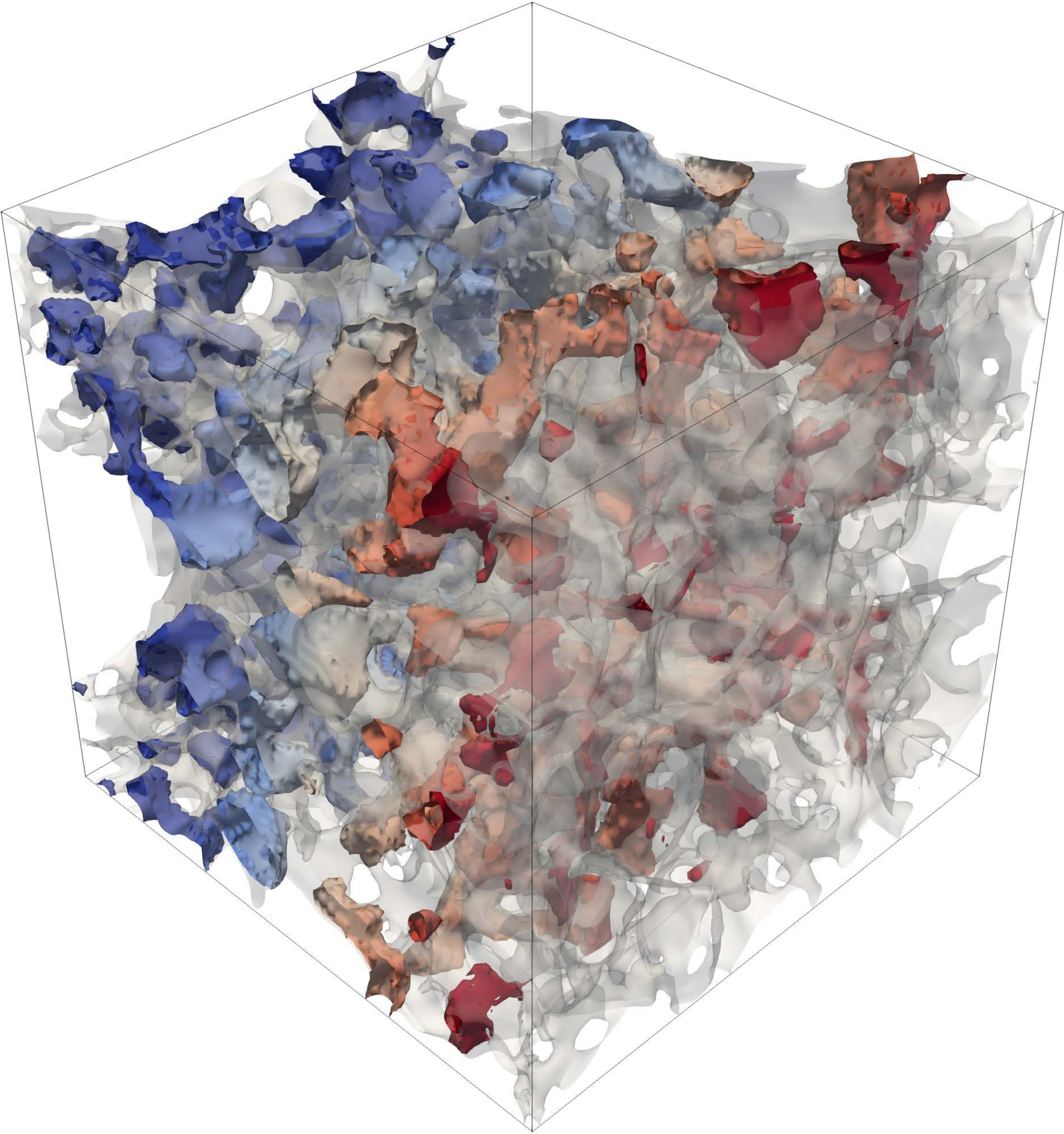} 
	\includegraphics[width=0.22\textwidth]{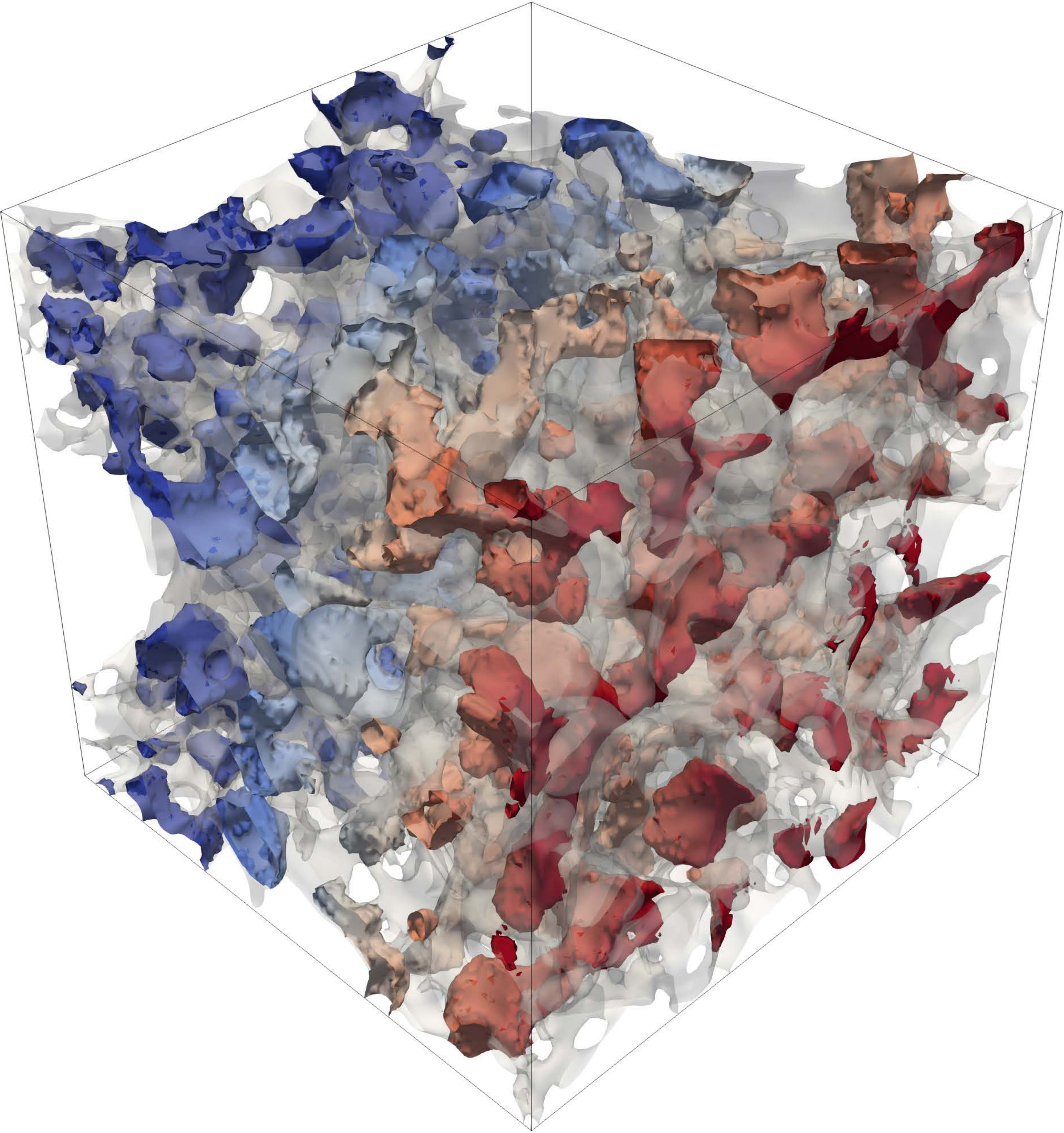}

	\includegraphics[width=0.22\textwidth]{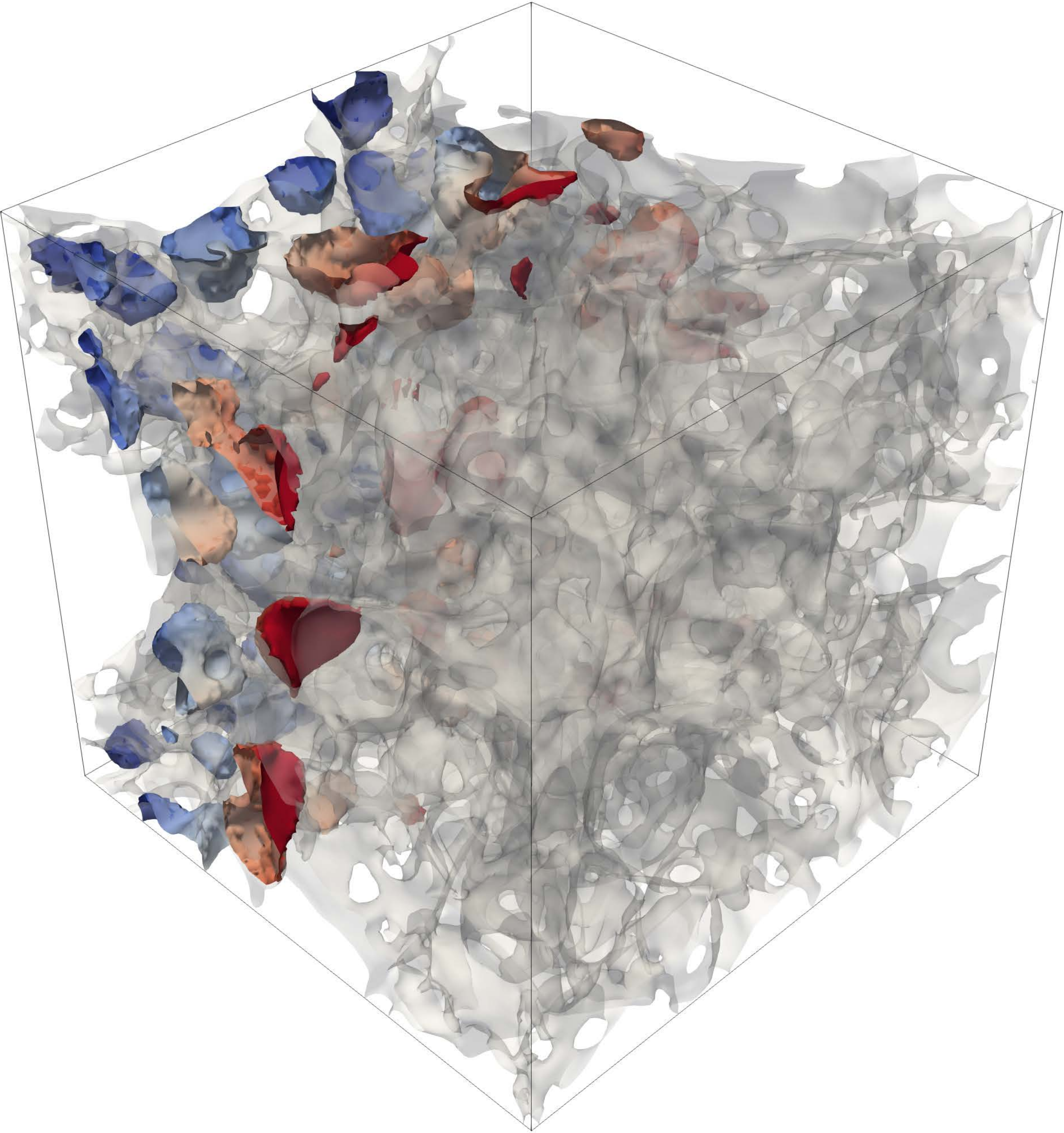} 
	\put(-115 ,60){(\textit{d})}
	\includegraphics[width=0.22\textwidth]{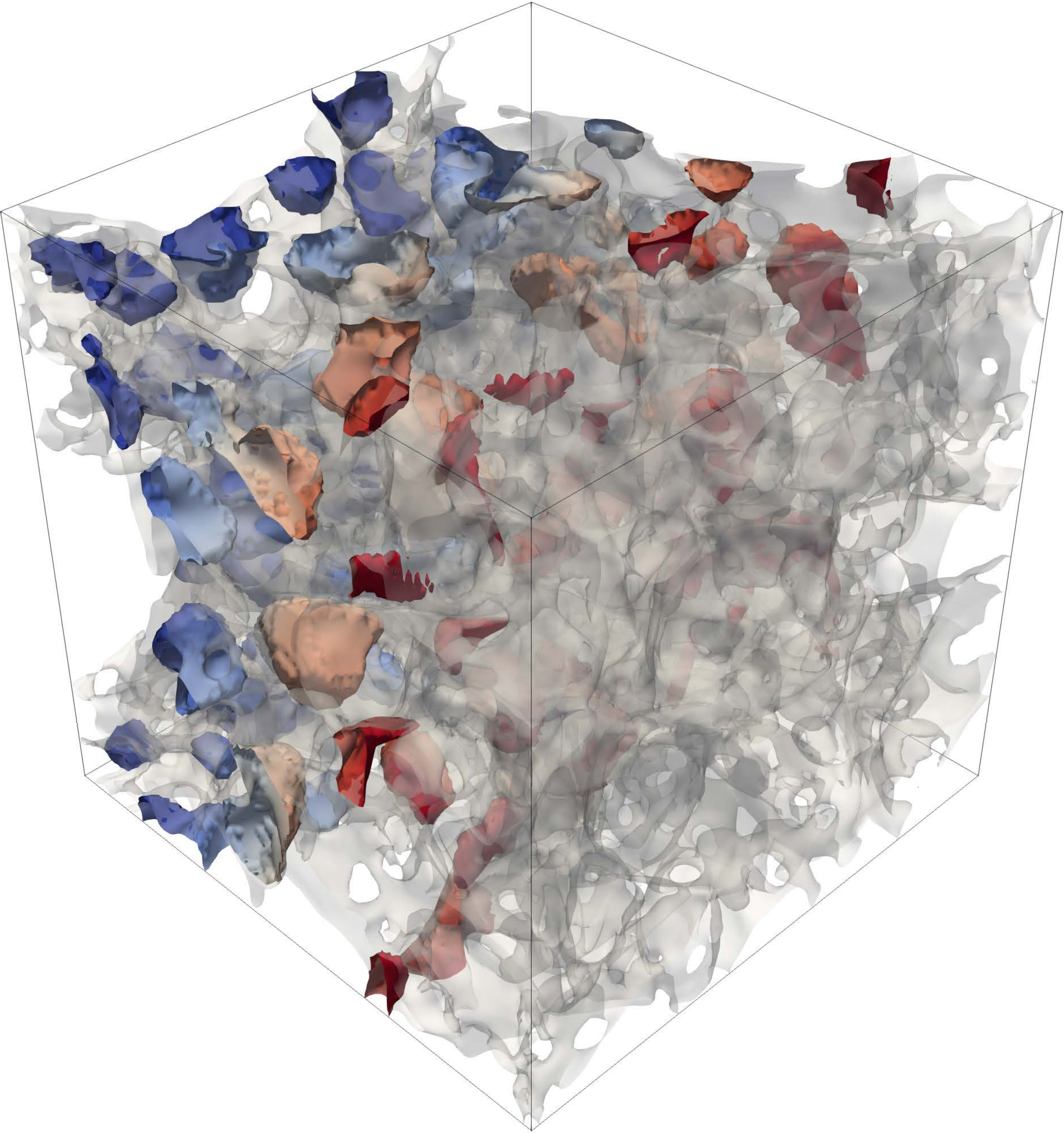}
	\includegraphics[width=0.22\textwidth]{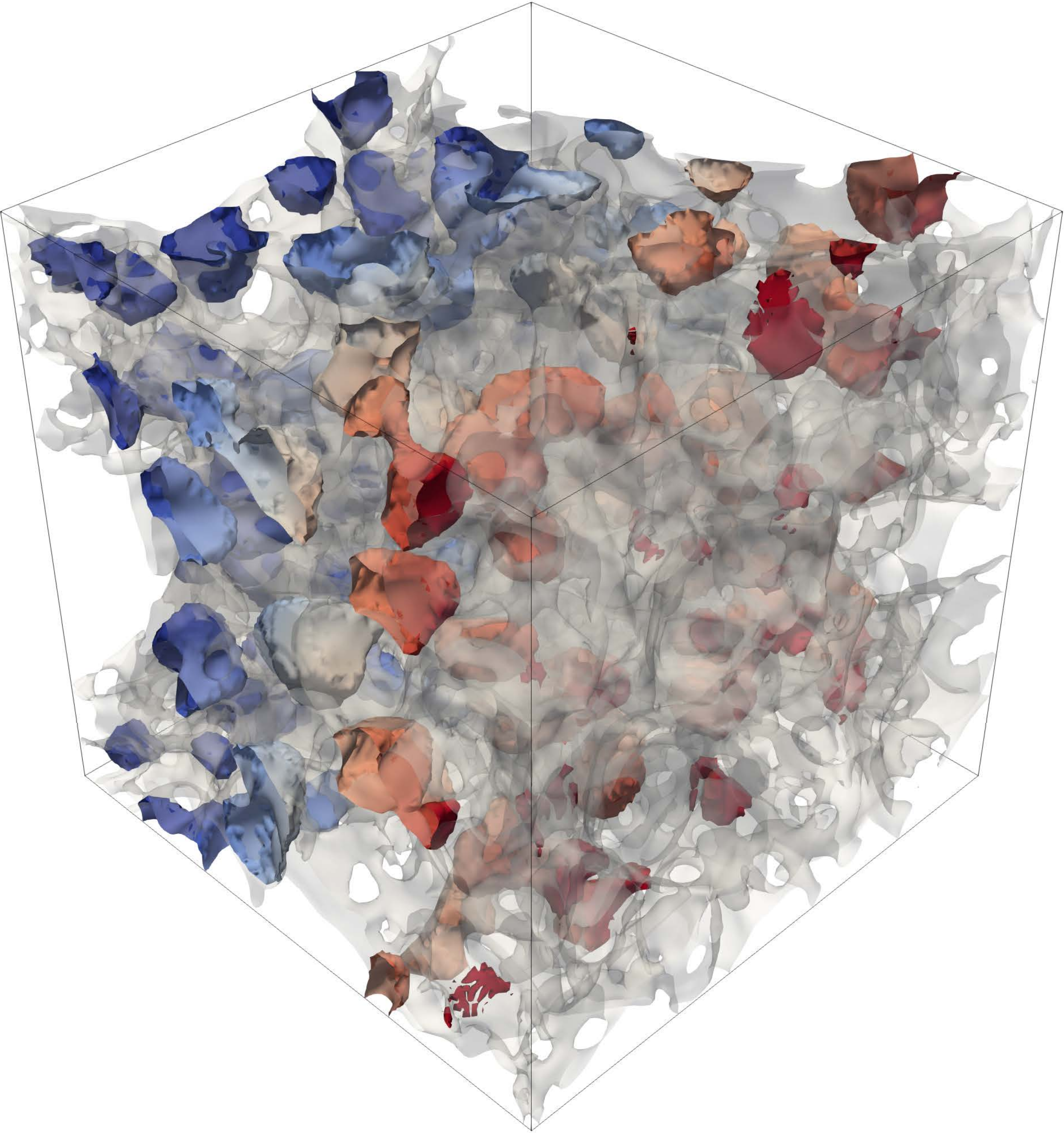} 
	\includegraphics[width=0.22\textwidth]{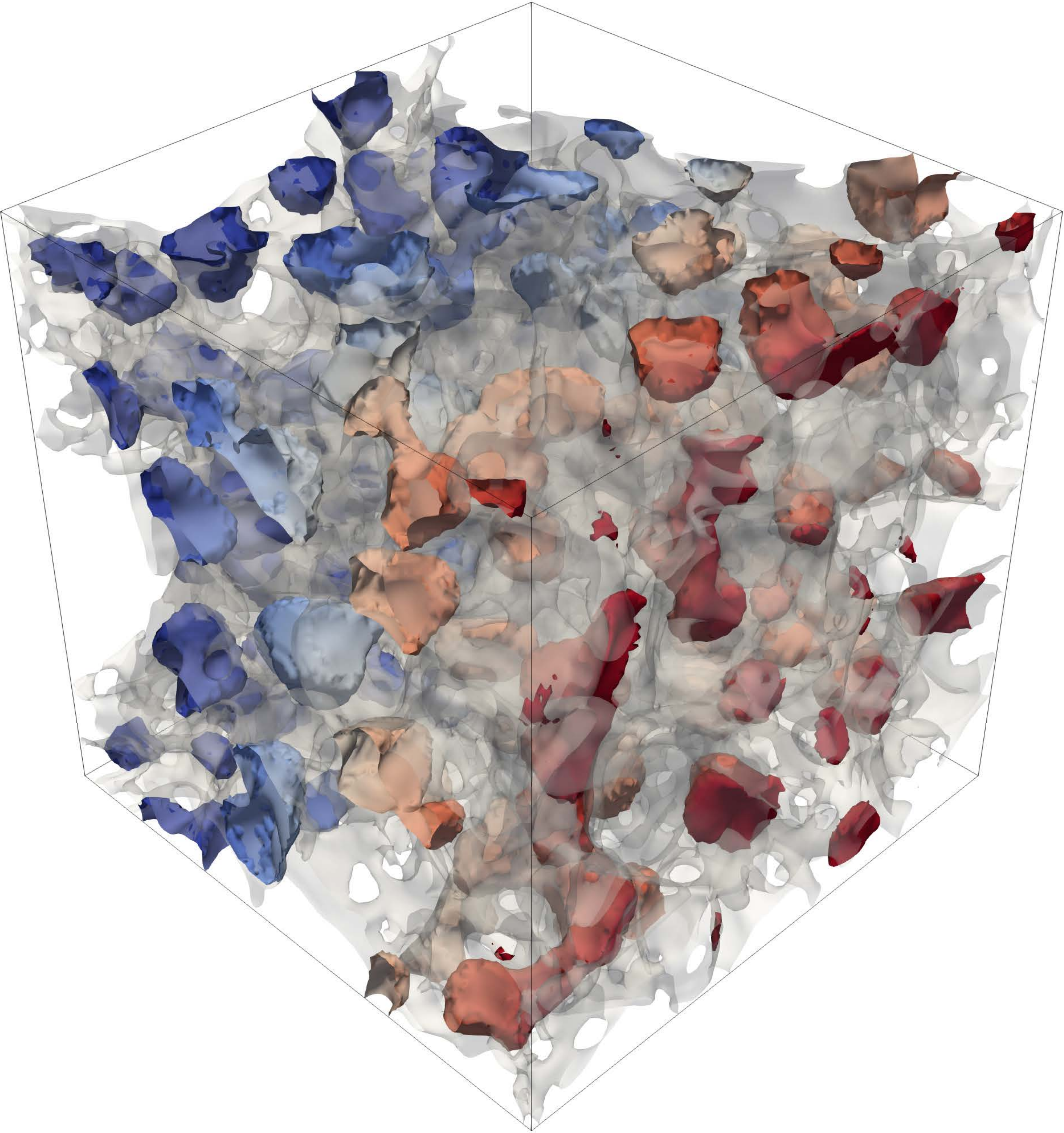}

	\includegraphics[width=0.22\textwidth]{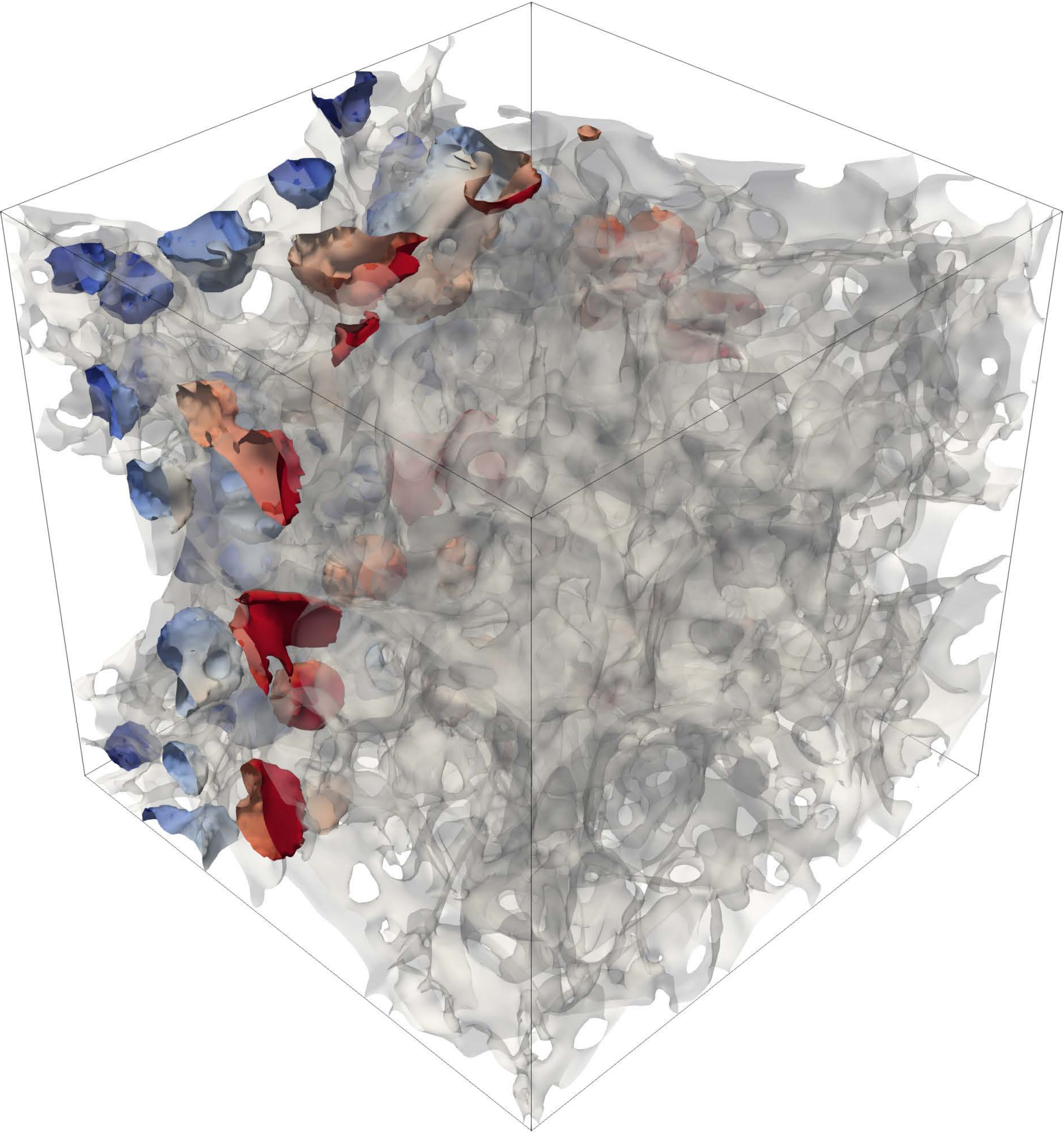} 
	\put(-115 ,60){(\textit{e})}
	\includegraphics[width=0.22\textwidth]{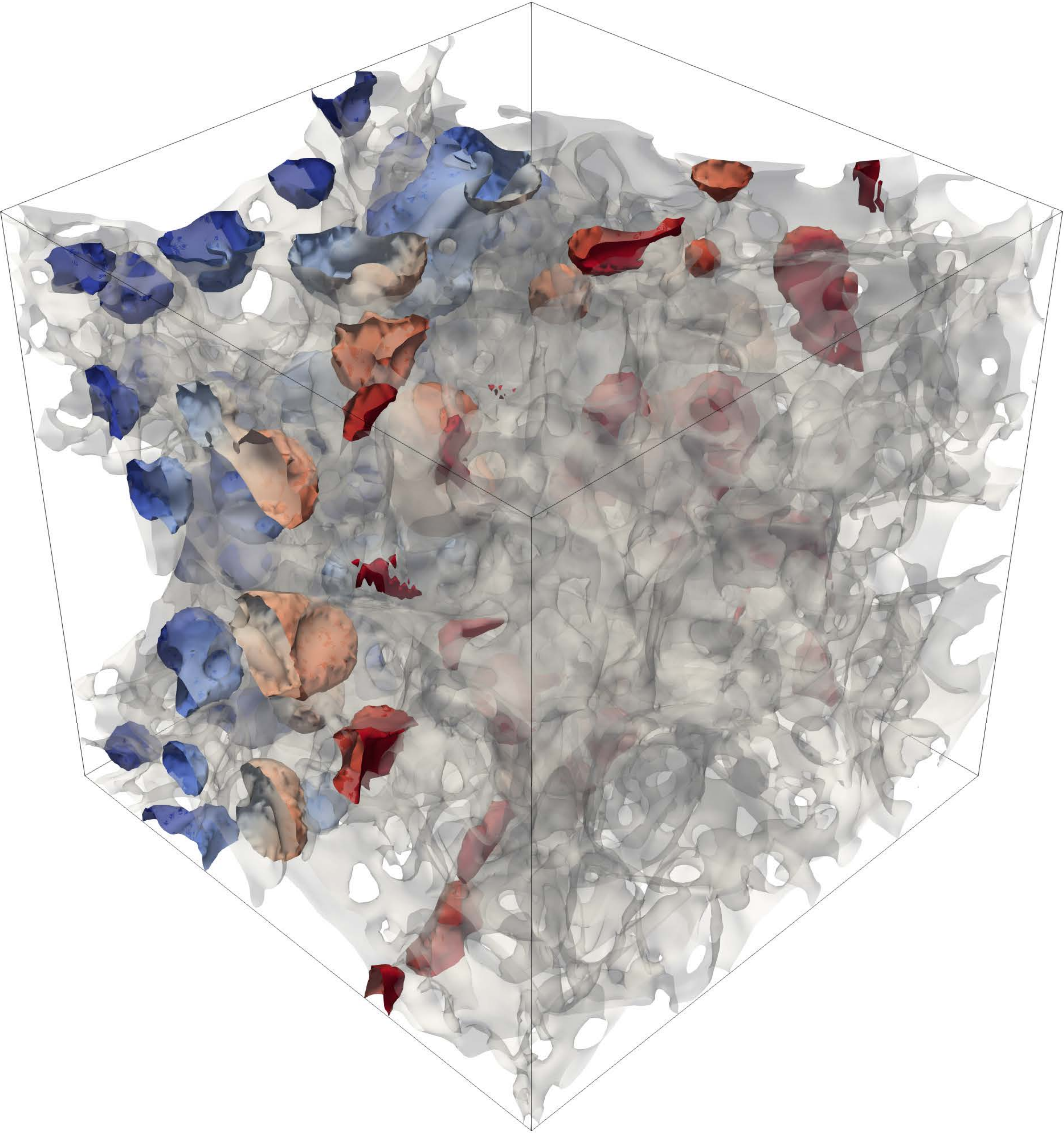}
	\includegraphics[width=0.22\textwidth]{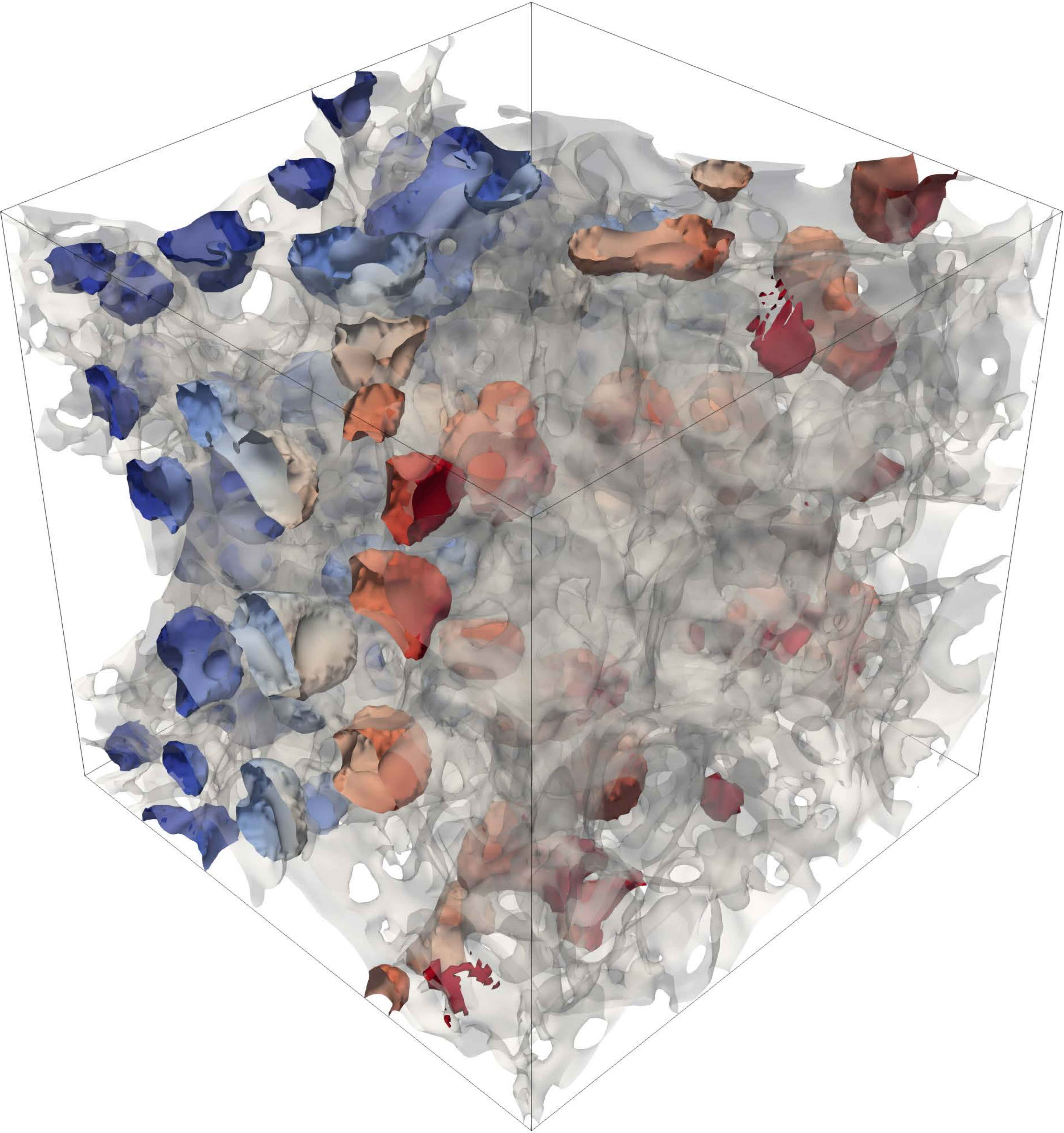} 
	\includegraphics[width=0.22\textwidth]{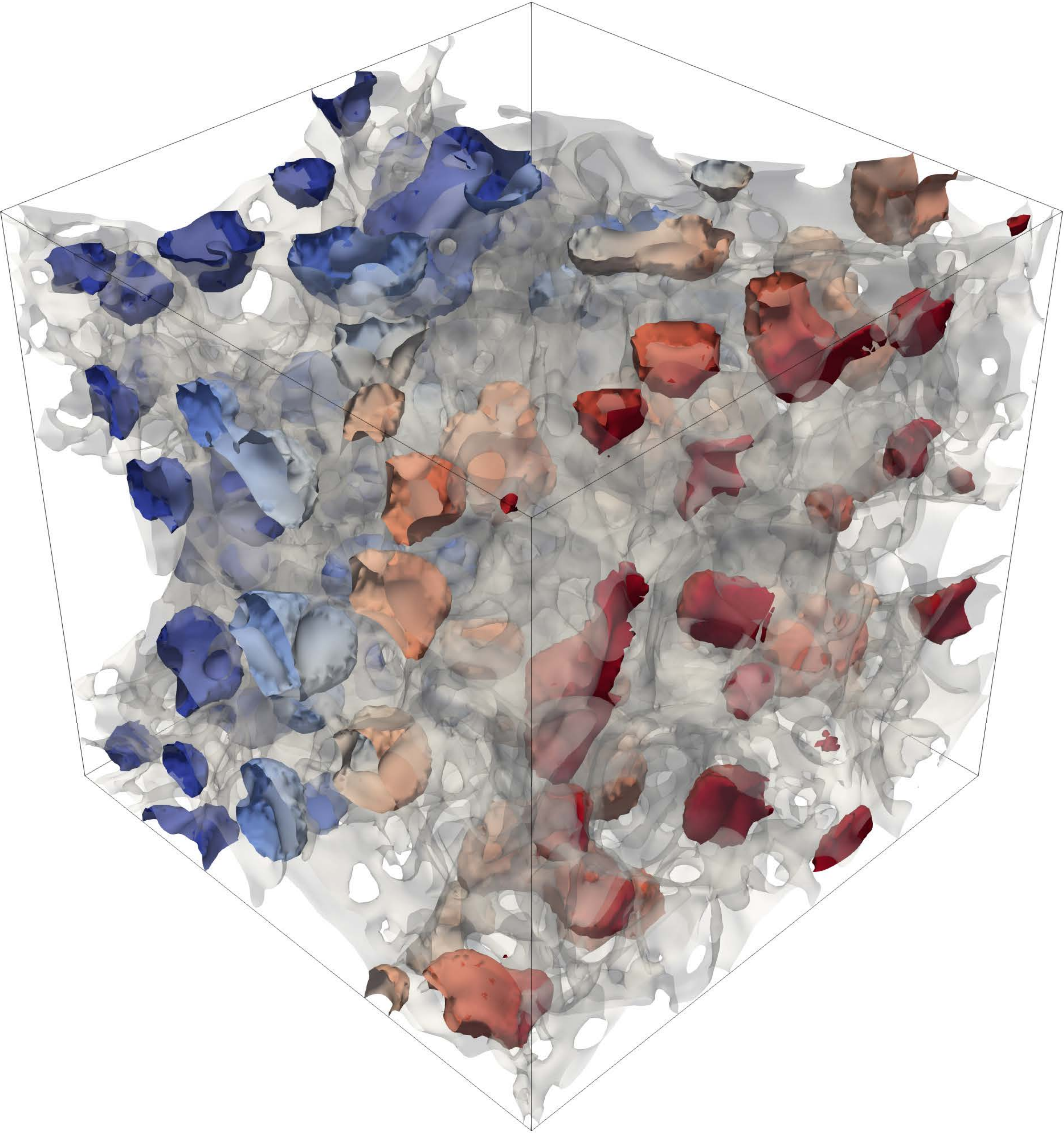} 	

	\caption{ The freezing fronts and temperature distributions under different contact angles: $\theta = 30^{\circ} $ (a), $\theta = 60^{\circ} $ (b), $\theta = 90^{\circ} $ (c), $\theta = 120^{\circ} $ (d), $\theta = 150^{\circ} $ (e), time increments from left to right}
	\label{fig12}
\end{figure}

Fig. \ref{fig13} shows the evolutions of solid fraction $f_s$ under different contact angles. Initially, the solid fractions of all cases are almost the same, this is because at this stage, the heat conduction is the dominant mode of heat transfer. With the increase of time ($Fo$), however, the freezing rate gradually decreases, indicating a slowdown in the advancement of the freezing front. This behavior is attributed to the initially steep temperature gradient, which promotes rapid freezing, followed by a reduced temperature gradient and increased thermal resistance that slow the freeing process. From this figure, one can also find that the freezing rate is also influenced by the wettability of solid surface. For the case with a small contact angle, the liquid is more prone to spread on the skeleton of porous medium, increasing the contact area and enhancing heat transfer. However, for the case with a large contact angle, the liquid usually distributes as the isolated droplets with reduced contact area, which limits thermal conduction and slows down the freezing process.

\section{Conclusions}
\label{sec5}

Based on phase-field theory and the enthalpy method, we develop a diffuse-interface model to study the multiphase freezing processes in complex geometries. This model can implicitly capture the evolution of the solid-liquid interface, and simultaneously, the effect of wettability can be incorporated into the energy functional, effectively overcoming the difficulty in directly implementing of wetting boundary condition on a complex boundary. In addition, through introducing a mass source term into the continuity equation, the model can quantitatively characterize the volume expansion or shrinkage caused by density difference during the freezing process. To solve the developed diffuse-interface model, an LB method is further proposed, and is tested  through several benchmark problems. For the conductive freezing problem with an infinite domain, the numerical results agree well with the theoretical solutions, demonstrating the accuracy of the enthalpy method in capturing the solid-liquid phase-change interface. The results of liquid column freezing show that the LB method can accurately represent phase change dynamics with the volume change. For the freezing process of droplets on both flat and curved cold substrates, the numerical results not only consistent with experimental data, but also demonstrate the effect of density ratio on the freezing morphology. For the case $\rho_s<\rho_l$ (volume expansion), a conical structure is formed at the top of the frozen droplet, while for the case $\rho_s>\rho_l$ (volume shrinkage), a flat surface morphology at the top of the frozen droplet is observed. The method is further extended to study the freezing processes in the fracture and unsaturated porous medium, and the results show that it can effectively capture the evolutions of the ice front and the gas-liquid interface, and also the wettability of solid surface and pore structure of porous medium have some significant influences on the freezing path. Finally, the present diffuse-interface method is expected to be an effective tool for complex freezing and solidification problems.

\section*{Acknowledgments}
This research was supported by the National Natural Science Foundation of China (Grant Nos. 123B2018 and 12501599), the Postdoctoral Fellowship Program of CPSF (Grant No. GZB20250714), China Postdoctoral Science Foundation (Grant No. 2025M773077), the Interdisciplinary Research Program of HUST (2024JCYJ001 and 2023JCYJ002), and the Fundamental Research Funds for the Central Universities, HUST (2024JYCXJJ016). The computation was completed on the HPC Platform of Huazhong University of Science and Technology.


\end{document}